\documentclass[useAMS,usenatbib]{mn2e}
\usepackage{graphicx}
\usepackage{url}
\usepackage{color}
\usepackage{bm}
\usepackage{amssymb}
\usepackage{amsbsy}
\usepackage{amsmath}
\usepackage{hyperref}
\usepackage{times}
\usepackage{rotating}
\usepackage{subfigure}

\newcommand\T{\rule{0pt}{2.6ex}}
\newcommand\B{\rule[-1.2ex]{0pt}{0pt}}

\hypersetup{
    unicode=true,                     
    linktocpage=true,
    pdftoolbar=true,                
   pdfmenubar=true,              
    pdffitwindow=true,           
    pdfstartview={FitH},        
    pdfhighlight={/I},
   hyperindex=true,
    pdftitle={My title},            
   pdfauthor={Jonathan Zwart},         
   pdfsubject={Subject},       
   pdfcreator={Creator},       
    pdfproducer={Producer}, 
    pdfkeywords={keywords}, 
    pdfnewwindow=true,      
   colorlinks=true,       
    linkcolor=red,          
    citecolor=blue,        
    filecolor=magenta,      
    urlcolor=cyan           
}

\newcommand{\me}{\mathrm{e}}
\newcommand{\dif}{\mathrm{d}}

\newcommand{\Ks}{\ensuremath{K_{\mathrm{s}}\,}}

\newcommand{\nest}{\textsc{MultiNest}}

 \title[Bayesian constraints on sub-$\mu$Jy source populations at
 1.4\,GHz]{Far beyond stacking: Fully bayesian constraints on
   sub-microJy radio source populations over the XMM-LSS--VIDEO field}

\author[Zwart, J.~T.~L.~et~al.]{Jonathan
  T.~L.~Zwart\thanks{jzwart@uwc.ac.za},$^{1,2}$  Mario Santos$^{1,3}$ and Matt J.~Jarvis$^{1,4}$
\\
$^1$Department of Physics \& Astronomy, University of the Western
Cape, Private Bag X17, Bellville, Cape Town 7535, South Africa\\
$^2$Astrophysics, Cosmology \& Gravity Centre, Department of Astronomy, 
University of Cape Town, Private Bag X3, Rondebosch 7701, South Africa\\
$^3$SKA South Africa, 3rd Floor, The Park, Park Road, Pinelands 7405, South Africa\\
$^4$Astrophysics, Department of Physics, Keble Road, Oxford OX1 3RH\\
}

\begin{document}
\date{Accepted ---. Received ---; in original form \today.}

\pagerange{\pageref{firstpage}--\pageref{lastpage}} \pubyear{2015}

\label{firstpage}

\maketitle

\begin{abstract}
  Measuring radio source counts is critical for characterizing new
  extragalactic populations, brings a wealth of science within reach
  and will inform forecasts for SKA and its pathfinders. Yet there is
  currently great debate (and few measurements) about the behaviour of
  the 1.4-GHz counts in the $\mu$Jy regime. One way to push the counts
  to these levels is via `stacking', the covariance of a map with a
  catalogue at higher resolution and (often) a different
  wavelength. For the first time, we cast stacking in a fully bayesian
  framework, applying it to (i) the SKADS simulation and (ii) VLA data
  stacked at the positions of sources from the VIDEO survey. In the
  former case, the algorithm recovers the counts correctly when
  applied to the catalogue, but is biased high when confusion comes
  into play. This needs to be accounted for in the analysis of data
  from any relatively-low-resolution SKA pathfinders. For the latter
  case, the observed radio source counts remain flat below the
  5-$\sigma$ level of 85\,$\mu$Jy as far as 40\,$\mu$Jy, then fall off
  earlier than the flux hinted at by the SKADS simulations and a
  recent $P(D)$ analysis (which is the only other measurement from the
  literature at these flux-density levels, itself extrapolated in
  frequency). Division into galaxy type via spectral-energy
  distribution reveals that normal spiral galaxies dominate the counts
  at these fluxes.
\end{abstract}


\section{Introduction}
\label{sec:intro}

Measurements of radio source counts provided some of the earliest
cosmological support for an expanding Universe, but today they can be
used to discover and characterize new extragalactic populations (see
e.g.~\citealt{AT20G,AT20G-opt,imogen2013,franzen2013}) and/or to shed
light on galaxy evolution in allowing (together with redshift
information) inference of star-formation rates and luminosity
functions (see e.g.~\citealt{dunne2009,karim2011,R+B2013,zwart2014}).
It is also critical to pin them down in advance of the new wave of
radio telescopes such as MeerKAT, ASKAP and SKA1: since confusion
noise (\citealt{scheuer57,condon74,condon1992,condon2002,condon2012})
depends directly on the source counts and a telescope's synthesized
beam, the \textit{level} of confusion noise is both a function of
these two \textit{and} informs the desired telescope configuration. It
is therefore essential to measure counts for both telescope design and
survey forecasting for SKA and its pathfinders (see
e.g.~\citealt{jz-skasci}). See also \cite{dezotti-review} for a review
of the counts at a range of radio frequencies.

There is currently considerable debate about the behaviour of 1.4-GHz
differential source counts at $\lesssim\mu$Jy levels, ARCADE2
\citep{arcade2,arcade2-2} finding them to be relatively flat (in the
Euclidean normalization, i.e.~flux density $S^{2.5}$) below
100$\mu$Jy, but with VLA 3-GHz measurements by \cite{condon2012}
hinting that they are rather steeper (roughly $\propto S^{-1.5}$ from
that work) at those flux-density levels. The work of \cite{om2008}
showed the need for careful corrections for source angular diameters
when assembling 1.4-GHz source counts. \cite{hjc2013} suggested the
considerable variation in the source counts at 100$\mu$Jy could be
attributed to sample variance. Indeed, a more-detailed analysis (see
below) of the \citeauthor{condon2012} data by \cite{vernstrom2014}
confirmed the steep counts down to sub-$\mu$Jy levels; their result is
still compatible with the \citeauthor{arcade2} data only with a new,
somewhat extreme population of very faint sources. Finally, the
state-of-the-art SKADS simulations \citep{wilman2008} also require the
source counts to be steep (and steepening) at the $\mu$Jy level. With
all this in mind, there is a clear need for more and complementary
measurements in this area, especially in the run up to SKA1 and as its
pathfinders come online.

Three main methods exist for determining source counts. In the first,
traditionally astronomers simply counted directly-detected sources in
flux bins above some well-motivated threshold (usually $5\times$ the
rms noise for the radio-source catalogue, but e.g.~\citealt{10C-II}
even used $4.62\times$ rms).

Second, \cite{scheuer57} showed that faint, unresolved sources buried
within the thermal noise could contribute a variance term from their
(positive) flux as well as from their (negative) synthesized-beam
sidelobes. Hence (see also above) measuring the noise allows one to
constrain source counts (a so-called probability of deflection, or
$P(D)$, analysis) if the confusion contribution is dominant.
\cite{hewish61} was the first to use Scheuer's blind method, and it
was most recently implemented (in the image plane) at 3~GHz by
\cite{condon2012} and \cite{vernstrom2014}, then extrapolated to
1.4\,GHz.

The third approach is `stacking'. One would like to push source-count
measurements as deep as possible, but a $P(D)$ analysis is not
possible unless a radio map is confusion-noise-limited. However, if
one knew in advance the positions of sources below the formal
detection threshold, for example from some deeper auxiliary catalogue,
one could constrain source counts below the detection threshold using
this prior information and accounting for that thermal noise. Such a
selection effect is a both a blessing and a curse: stacking allows for
a robust `source-frame' definition and subsequent subdivision of the
sample, but cannot ever capture properties of a population defined by
an `observer-frame' blind analysis such as $P(D)$.
The method has been applied in a number of different settings, e.g.~in
polarization \citep{stil2014}, in the sub-mm \citep{patanchon2009} and
at 1.4\,GHz \citep{seymour2008,atlas2,atlas2-2}.

\cite{blast2009} formally define stacking as taking the covariance of
a map with a catalogue (at any two wavebands), but in fact it has many
flavours in the literature as pointed out by \cite{zwart2014}. In the
past \citep{dunne2009,karim2011,zwart2014} some single statistic
(usually an average such as the mean or median) has been used to
summarize the properties of radio pixels stacked at the positions from
the auxiliary catalogue, allowing calculation of, for example,
star-formation rates in mass-redshift bins. The precision on any
inferred parameters scales as $N_{\mathrm{sources}}^{-1/2}$ \citep{R+B2013}, so
for (say) 40,000 sources a statistical factor of 200 in sensitivity
below the survey threshold could in theory be achieved.

However, the data afford more information than what is encoded in a
single summary statistic (and careful attention must be paid to
potential biases in those analyses; see \citealt{jz-skasci}). For
example, \cite{bourne2012} identify three potential sources of bias in
median stacking: the flux limits and shape of the `true' underlying
flux distribution, and the amplitude of the thermal noise. Rather than
being viewed as biases, such extra information in the observed data
can be exploited in order to infer the parameters of some underlying
physical model.

This was demonstrated by \cite{ketron2014} who constrained 1.4-GHz
source counts down to $\sigma/10$ from FIRST maps \citep{first}
stacked, at positions from deeper COSMOS data \citep{cosmos}, by
fitting a power-law model to the noisy map-extracted flux
distribution. Using a similar parametric method, \cite{R+B2013}
included redshift information in order to measure the evolution of the
1.4-GHz luminosity function for near-infrared-selected galaxies, which
they called `parametric stack-fitting'.

In this work we extend the algorithm of \cite{ketron2014} in the
following ways: (i) We cast the problem in a fully bayesian framework
rather than in a purely maximum-likelihood one; (ii) in our framework
we consider the use of the bayesian evidence for selecting between
different source-count models; we demonstrate the algorithm on the
SKADS simulations; and (iv) we apply the algorithm to a deep ($\Ks<23.5$)
mass-selected data sample from the VIDEO survey \citep{jarvis2013}.

In section~\ref{sec:bayes} we describe our bayesian framework,
including the models, priors and likelihoods we use. We give some
details of simulations undertaken in section~\ref{sec:tests:sims},
with a description of the near-infrared and radio observations and
data in section~\ref{sec:obs}. We present our results and discussion
in section~\ref{sec:results} before concluding in
section~\ref{sec:conclude}.

We assume radio spectral indices $\alpha$ are such that
$S_{\nu}\propto\nu^{\alpha}$ for a source of flux density $S_{\nu}$ at
frequency $\nu$. All coordinates are epoch J2000. Magnitudes are in
the AB system. We assume a $\Lambda$CDM `concordance' cosmology
throughout, with $\Omega_{\mathrm{m}}=0.3$, $\Omega_{\Lambda}=0.7$ and
$H_0= 70\,\mathrm{km\,s^{-1}}\,\mathrm{Mpc}^{-1}$ \citep{wmap7params}.

\section{Bayesian Framework}
\label{sec:bayes}

\subsection{Bayes' Theorem}
\label{sec:bayes:theorem}

Bayesian analyses of astronomical data sets have become commonplace in
the last ten years (see e.g.~\citealt{mcadam}, \citealt{feroz-kepler},
\citealt{ami-one}, \citealt{biro}) because of their many advantages.
The targeted posterior probability distribution
$\mathcal{P}\left(\mathbf{\Theta}|\mathrm{\mathbf{D}},H\right)$ of the
values of the parameters $\mathbf{\Theta}$, given the available data
$\mathrm{\mathbf{D}}$ and a model $H$ (the model is a hypothesis plus
any assumptions), comes from Bayes' theorem:

\begin{equation} 
\label{eqn:bayes}
\mathcal{P}\left(\mathbf{\Theta}|\mathrm{\mathbf{D}},H\right)
= \frac{
\mathcal{L}\left(\mathrm{\mathbf{D}}|\mathbf{\Theta},H\right)
\mathit{\Pi}\left(\mathbf{\Theta}| H\right)}
{\mathcal{Z}\left(\mathrm{\mathbf{D}}| H\right)}.
\end{equation} 

\noindent In the numerator, the likelihood
$\mathcal{L}\left(\mathrm{\mathbf{D}}|\mathbf{\Theta},H\right)$,
i.e.~the probability distribution of the data given parameter values
and a model, encapsulates the experimental constraints. The prior
$\mathit{\Pi}\left(\mathbf{\Theta}| H\right)$ records prior knowledge
of or prejudices about the values of the parameters. The bayesian
evidence, $\mathcal{Z}\left(\mathrm{\mathbf{D}}| H\right)$, is the
integral of
$\mathcal{L}\left(\mathrm{\mathbf{D}}|\mathbf{\Theta},H\right)
\mathit{\Pi}\left(\mathbf{\Theta}| H\right)$
over all $\mathbf{\Theta}$. Na\"{\i}vely it normalizes the posterior
in parameter space; crucially it facilitates selection between
different models when their evidences are compared quantitatively
(Occam's razor; see e.g.~\citealt{mackay03}).

\subsection{Sampling}
\label{sec:bayes:sampling}

Carrying out the evidence integrations and sampling the parameter
space has not in the past been easy and was often slow.
Nested sampling \citep{skilling04} was invented specifically to reduce
the computational cost of evidence calculations, since no posterior
sample (each of which is obtained `for free') is ever wasted. However,
the integrations are exponential in the number of model parameters,
typically limiting that number to O($10^2$) on current machines. A
particularly robust and efficient implementation of nested sampling is
\nest\ \citep{feroz08,feroz09,pymultinest}, which permits sampling
from posteriors that are multimodal and/or unusually shaped. Posterior
distributions are represented by full distributions rather than a
summary mean/median value and a (perhaps covariant) uncertainty, since
this represents the total inference about the problem at hand, and
error propagation is fully automatic.

In what follows we used a python interface \citep{pymultinest} to
\nest, deployed in parallel on (typically) 96 processors. Because of
the unusually-shaped posterior distributions, there were often as many
as O(100,000) likelihood evaluations. However, the resources required
for the algorithm are not heavy on either memory or disk space: it is
the number of parallel processors that controls the (rejection) sampling
efficiency and hence total wall execution time.

\subsection{Models considered here}
\label{sec:bayes:models}

We treat four source-count models here, each encoding some functional
form with a number of parameters and incorporating priors on each of
those parameters, plus any assumptions. The evidence will be employed
to select between several piecewise-linearly-interpolated power-law
models (A, B, C and D)
by way of example, but any parametric model could be used be it a
modified power law, a polynomial or a pole/node-based model (see
e.g.~\citealt{optimal-binning}, \citealt{vernstrom2014}).

\subsubsection{Model A}
\label{sec:bayes:models:single}

For the single-power law, we have simply that

\begin{equation}
\label{eqn:spl}
\frac{\dif N}{\dif S}
\left(C,\alpha,S_{min},S_{max}\right)
=
\begin{cases}
   CS^{\alpha} & S_{min}<S<S_{max} \\
    0      & \mathrm{otherwise} \\
\end{cases},
\end{equation}

\noindent where $\frac{\dif N}{\dif S}$ is the differential count for
$N$ sources in the flux interval $\left[S,S+\dif S\right]$
($\mathrm{Jy}$). $C$ is a normalization constant
($\mathrm{sr}^{-1}\mathrm{Jy}^{-1}$), $\alpha$ is a slope, and the
model is set to zero outside the lower and upper limits $S_{min}$ and
$S_{max}$. We denote the parameter vector
$\mathbf{\Theta_A}=\{C,\alpha,S_{\mathrm{min}},S_{\mathrm{max}}\}$.

\subsubsection{Models B, C and D}
\label{sec:bayes:models:double}

Model B extends Model A with a second power law (two extra
parameters), i.e.~it incorporates a second slope $\beta$ with a break
at $S_0$ ($S_{min}<S_0<S_{max}$):

\begin{multline}
\label{eqn:tpl}
\frac{\dif N}{\dif S}
\left(C,\alpha,\beta,S_0,S_{min},S_{max}\right)\\
=
\begin{cases}
CS^{\alpha} & S_{min}<S<S_0\\
C\,S_0^{\alpha-\beta\,}S^{\beta} & S_0<S<S_{max}\\
    0      & \mathrm{otherwise}\\
\end{cases},
\end{multline}



\noindent with a parameter vector
$\mathbf{\Theta_B}=\{C,\alpha,\beta,S_0,S_{\mathrm{min}},S_{\mathrm{max}}\}$. Model
C incorporates another break in the power law, so
$\mathbf{\Theta_C}=\{C,\alpha,\beta,\gamma,S_0,S_1,S_{\mathrm{min}},S_{\mathrm{max}}\}$,
and Model D, a third break,
i.e.~$\mathbf{\Theta_D}=\{C,\alpha,\beta,\gamma,\delta,S_0,S_1,S_2,S_{\mathrm{min}},S_{\mathrm{max}}\}$. Priors
on the different model parameters are discussed in section
\ref{sec:bayes:priors}.

\subsubsection{Models A\arcmin, B\arcmin, C\arcmin\ and D\arcmin}
\label{sec:bayes:models:float-noise}

\noindent In section \ref{sec:results} we investigate some scenarios
where we allow the flux-measurement noise $\sigma_n$ (see section
\ref{sec:bayes:likelihoods}) to be a free parameter of the fit. We
denote these models A\arcmin, B\arcmin, C\arcmin\ and D\arcmin\
corresponding to their fixed-noise counterparts A, B, C and D
respectively.

\subsection{Priors}
\label{sec:bayes:priors}

It is straightforward to vary the priors
$\mathit{\Pi}\left(\mathbf{\Theta}| H\right)$ on parameters
$\mathbf{\Theta}$; the ones we adopt are shown in
Table~\ref{table:priors}. In particular the power-law normalization
$C$, as a scale parameter, is given a logarithmic prior, and any
power-law breaks are free to vary their positions.

We assume all model hypotheses to be equally likely \textit{ab
  initio}, which is equivalent to considering solely the Bayes factor
$\mathcal{Z}_1\left(\mathrm{\mathbf{D}}|
  H_1\right)/\mathcal{Z}_2\left(\mathrm{\mathbf{D}}| H_2\right)$
when comparing models.

\begin{table}
\centering
\caption{Priors $\mathit{\Pi}\left(\mathbf{\Theta}| H\right)$ adopted here.\label{table:priors}}
\begin{tabular}{ll}
\hline
Parameter \T\B & Prior  \\
\hline
$C/\mathrm{sr^{-1}Jy^{-1}}$ \T\B & log-uniform $\in \left[10^{-5},10^7\right]$ \\
$\alpha,\beta,\gamma,\delta$ \B & uniform $\in \left[-2.5,-0.1\right]$ \\
$S_{min}/\mu$Jy \B & uniform $\in \left[0.01,20.0\right]$ \\
$S_{max}/\mu$Jy \B & uniform $\in \left[20.0,100.0\right]$ \\
$S_{0,1,2}$ \B & uniform $\in \left[S_{min},S_{max}\right]$ \\
$S_{0,1,2}$ \B & further require $S_0<S_1<S_2$ \\
$\sigma_n$ \B & $\delta\left(\sigma_{\mathrm{survey}}\right)$, or \\
\B & uniform $\in \left[0.5,2.0\right]\sigma_{\mathrm{survey}}$ \\

\hline
\end{tabular}
\end{table}

\subsection{Likelihood Function}
\label{sec:bayes:likelihoods}

In what follows we drop an explicit dependence on $\mathcal{H}$ but do
note our assumptions along the way. We cannot use a simple gaussian
($\chi^2$) likelihood function because the measurement uncertainties
are not themselves gaussian: since we are dealing with binned data we
adopt a poisson likelihood. For the $i^{th}$ bin containing $k_i$
objects, the corresponding likelihood
$\mathcal{L}_{i}\left(k_i|\pmb{\theta}\right)$ is

\begin{equation}
\mathcal{L}_{i}\left(k_i|\pmb{\theta}\right)
=
\frac{I_i^{k_i}\me^{-I_i}}{k_i!},
\end{equation}

\noindent where

\begin{equation}
\label{eqn:ii}
I_i=
\int_{S_{min}}^{S_{max}}
\dif S \frac{\dif N(S)}{\dif S}
\int_{S_{m_i}}^{S_{m_i}+\Delta S_{m_i}}
\dif S_m
\frac{1}{\sigma_{n}\sqrt{2\pi}}
\me^{-\frac{\left(S-S_m\right)^2}{2\sigma_n^2}}.
\end{equation}

\noindent The second integral accounts for the gaussian map
noise. $I_i$ is, in a sense, a mock realisation of the count in a
particular bin from a specific set of parameter values, to be compared
with the real values encoded by $k$. Folded into $I_i$ are the
flux-measurement noise $\sigma_n$ (assumed here to be the same for all
objects, i.e.~constant across the field, $\sigma_j=\sigma_n\forall j$
sources) and the bin limits $\left[S_{m_i} ,S_{m_i}+\Delta
  S_{m_i}\right]$. $S$ and $S_m$ are respectively the measured and
noise-free fluxes for the sources in a given bin, related by
$S\sim\mathcal{N}\left(S_m,\sigma_n^2\right)$. Carrying out the second
integration in equation \ref{eqn:ii} gives

\begin{multline}
\label{eqn:iii}
I_i=
\int_{S_{min}}^{S_{max}}
\dif S \frac{\dif N(S)}{\dif S}
\\
\frac{1}{2}
\left\{\mathrm{erf}\left(\frac{S-S_{m_i}}{\sigma_n\sqrt{2}}\right)
- \mathrm{erf}\left(\frac{S-(S_{m_i}+\Delta S_{m_i})}{\sigma_n\sqrt{2}}\right)\right\}.
\end{multline}

\noindent Assuming independent bin entries, the likelihoods multiply
(the log-likelihoods adding) to give the total likelihood for the
$n_{bins}$ bins:

\begin{equation}
\label{eqn:lhood-tot-bins}
\mathcal{L}\left(\mathbf{k}|\pmb{\theta}\right)
=\prod_{i=1}^{n_{bins}} \mathcal{L}_{i}\left(k_i|\pmb{\theta}\right).
\end{equation}

\noindent
When combined with the priors from Table~\ref{table:priors}, inserting
equation~\ref{eqn:lhood-tot-bins} into equation~\ref{eqn:bayes} yields
the posterior probability distribution (for parameter estimation) and
the bayesian evidence (for model selection).

As a final point, it is possible to jointly infer an assumed-constant
measurement noise, including any contribution from confusion noise, at
the same time as the source-count model parameters. Hence the
source-count model could simultaneously be extended to even fainter
fluxes, as well as to the inclusion of brighter sources \textit{above}
the detection threshold.

\vspace{2\baselineskip}
\noindent
With the formalism for the priors and likelihoods in place, we have
all of the ingredients we need to calculate posterior probability
distributions and the evidences for each model. We now turn our
attention to testing the algorithm on simulations.

\section{Simulations}
\label{sec:tests:sims}

The Square Kilometre Array Design Studies SKA Simulated Skies
(SKADS--S$^3$) simulation \citep{wilman2008,skads2} is a
semi-empirical model of the extragalactic radio-continuum sky covering
an area of 400\,deg$^2$, from which we extracted a 1-deg$^2$ catalogue
at 1.4\,GHz for the purposes of testing our method. The simulation,
which is the most recent available, incorporates both large- and
small-scale clustering and has a flux limit of 10\,nJy. We undertook
several tests as follows.

\begin{figure*}
\centering
\includegraphics[width=1.0\textwidth,origin=br,angle=0]{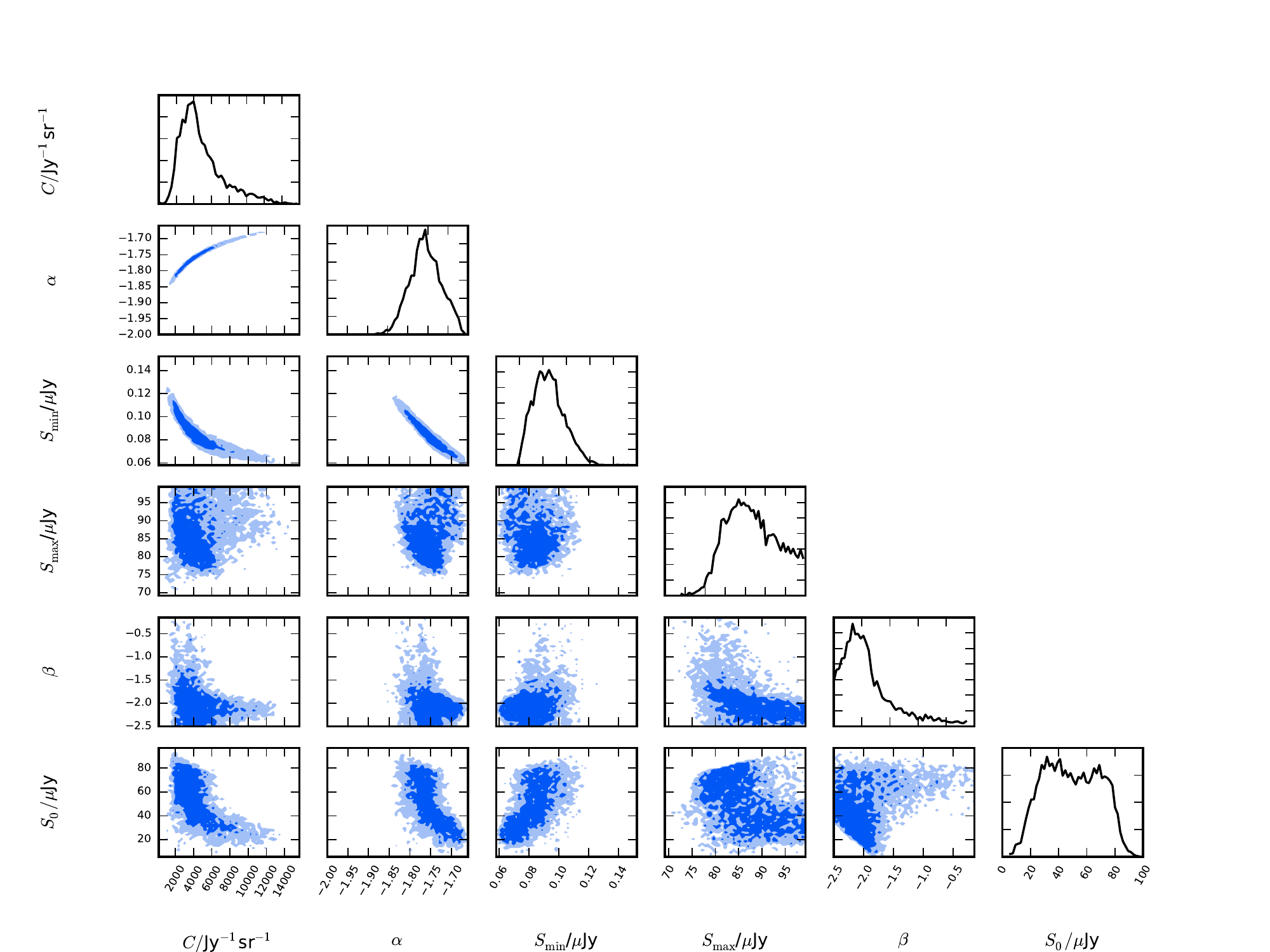}
\caption{Posterior probability distribution for the (winning)
  single-break model B fitted to the full SKADS mock catalogue data.
  The 68 and 95~per~cent confidence limits are respectively indicated by the dark and light
  shaded regions.\label{fig:skads-tri1}}
\end{figure*}

First, we took all $\approx$\,375,000 of the noise-free 1.4-GHz fluxes
($<85\mu$Jy, i.e.~the 5-$\sigma$ limit of our radio map --- see
section \ref{sec:obs:radio}) and added gaussian random noise of
standard deviation equal to the modal VLA map 1-$\sigma$ noise
(i.e.~16.2\,$\mu$Jy). We divided the resulting mock catalogue into 38
bins, $-68<S/\mu$Jy $<85$, with the lower edge of the lowest bin being
the flux of the faintest source in the noisy catalogue. Fitting each
of models A, B, C and D in turn, we found that model B (one break)
returned the highest evidence (Table
\ref{table:skads-evs}). Figure~\ref{fig:skads-tri1} is the posterior
probability distribution of the parameters for that model.

A bayesian framework allows us to explore the full posterior, rather
than simply determining the position of its peak, even though the
distribution is highly non-gaussian. For example, there is a strong
degeneracy between $C$ and $\alpha$; there is a hard diagonal edge in
the positions of adjacent power-law breaks because of the constraint
that e.g.~$S_0<S_1<S_2$; alternate power laws tend to be
anti-correlated in their parameters; and overall there is little
symmetry in the posterior. A fully bayesian analysis such as ours is
essential because so many of the assumptions of a maximum-likelihood
approach have broken down. The bayesian evidence is in turn the
appropriate tool to be used for model selection in place of a
$\Delta\chi^2$ --- an estimator that would not have told the whole
story.

\begin{figure}
\centering
\includegraphics[width=8cm,origin=br,angle=0]{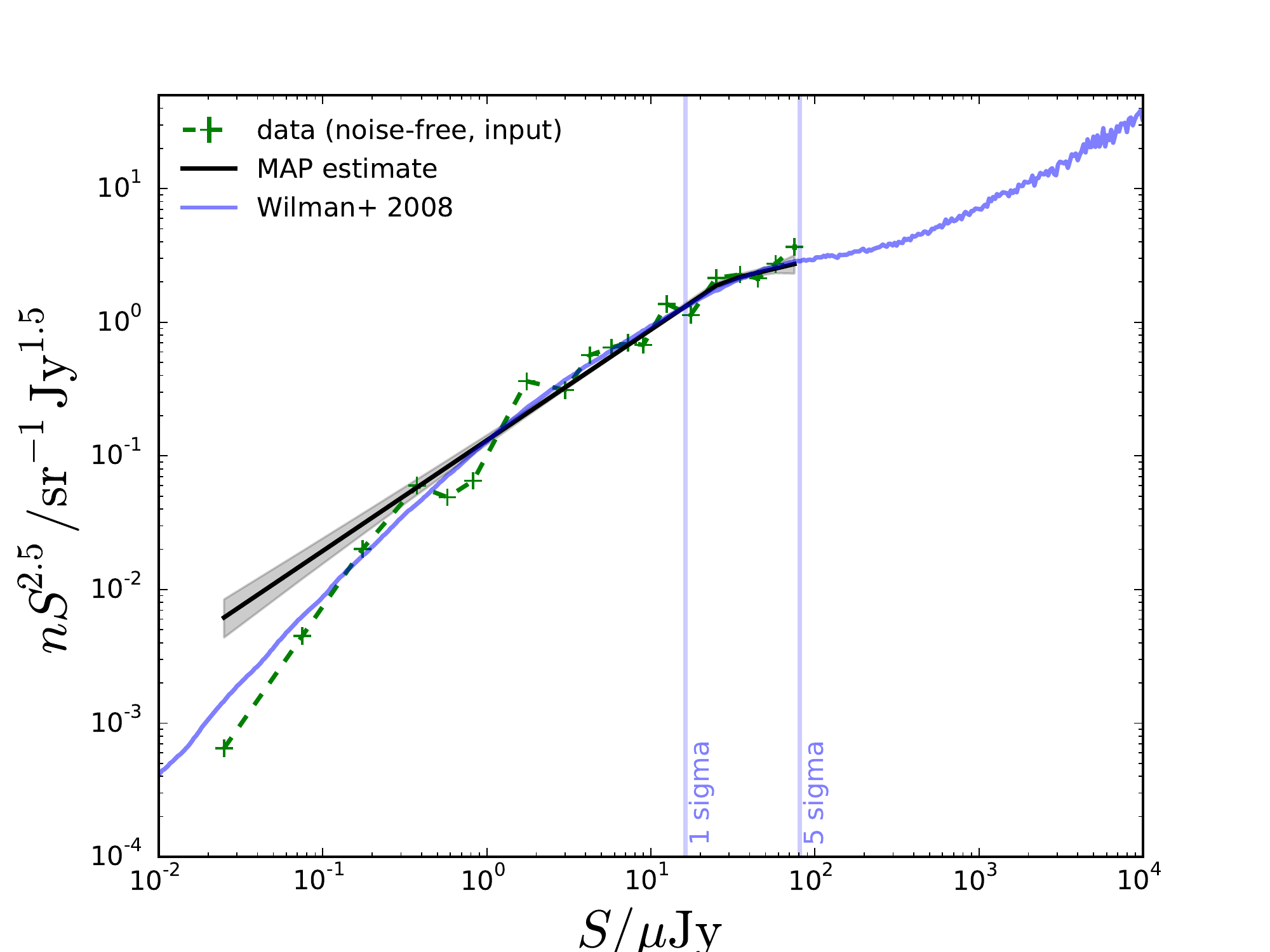}
\caption[]{Source counts reconstructed from the SKADS mock
  catalogue. The underlying, noise-free realisation of the counts for
  the 1-deg$^2$ catalogue is indicated by the green dashed line. The real, noisy
  catalogue data are shown as a black dashed line. The black line and
  grey shaded region show the maximum \textit{a posteriori} (MAP) and
  68-per-cent confidence region of the distributions of models reconstructed from
  every sample from the posterior in Figure~\ref{fig:skads-tri1}. The
  vertical blue lines are at 1$\sigma$ and 5$\sigma$.\label{fig:skads-recon1}}
\end{figure}

We have deliberately not tabulated maximum \textit{a posteriori}
(MAP), maximum-likelihood or `best-fit' parameter values because even
the 1-D marginalized posteriors are misleading in their masking of the
intricate characteristics of the full posterior probability
distribution. Point estimates such as the maximum-likelihood and MAP
are subject to noise and hence are not very representative of the full
posterior. Rather, we reconstruct source counts directly from the
ensemble of posterior samples, as follows.
Figure~\ref{fig:skads-recon1} shows the source-count reconstruction
based on (the winning) model B. For each sample from the posterior, we
have generated a corresponding source-count reconstruction evaluated
at each flux-bin centre; the uncertainties are defined as the
68-per-cent confidence region around the median. We have recentred
those uncertainties on the MAP reconstruction. Note that the
uncertainties are underestimates because they do not account for
correlations in data space: for example, high signal-to-noise data
(e.g.~at the bright end) will generally drive the uncertainties in the
power law (which is fitted over the full flux range, making the bins
not independent), but the error bars inflate at the faint end where
the signal-to-noise is low (see below).

The reconstructed count is consistent with the underlying SKADS count
to below 1\,$\mu$Jy and plausibly as low as 0.3\,$\mu$Jy (given also
the input realisation of that underlying count; green dashed line).
The counts are overestimated at the lowest fluxes, where we note that
there is little signal-to-noise. In fact the algorithm exhibits
$1/\sqrt{N}$ behaviour: for $N=375,000$, one might hope to reach a
factor of $\approx\sqrt{N}$ below the 5-$\sigma$ survey threshold of
$85\mu$Jy, i.e.~0.1\,$\mu$Jy, which is indeed roughly what we see.

\begin{table}
\centering
\caption{Evidences $\Delta\mathcal{Z}$ for models A--D applied to the
  full (1-deg$^2$) SKADS catalogue, relative to model A. The preferred model is B.
  \label{table:skads-evs}}
\begin{tabular}{lcc}
\hline
Model \T\B & $\Delta\log_{\mathrm{e}}\mathcal{Z}$ & Odds ratio $\mathcal{Z}/\mathcal{Z}_A$\\
\hline
A \T & 0.0 & 1 \\
B & \textbf{0.87} $\pm$ \textbf{0.20} & 2.39:1 \\
C & 0.83 $\pm$ 0.21 & 2.29:1 \\
D \B & 0.34 $\pm$ 0.21 & 1.40:1 \\
\hline
\end{tabular}
\end{table}

\begin{figure*}
\centering
\includegraphics[width=8.5cm,origin=br,angle=0]{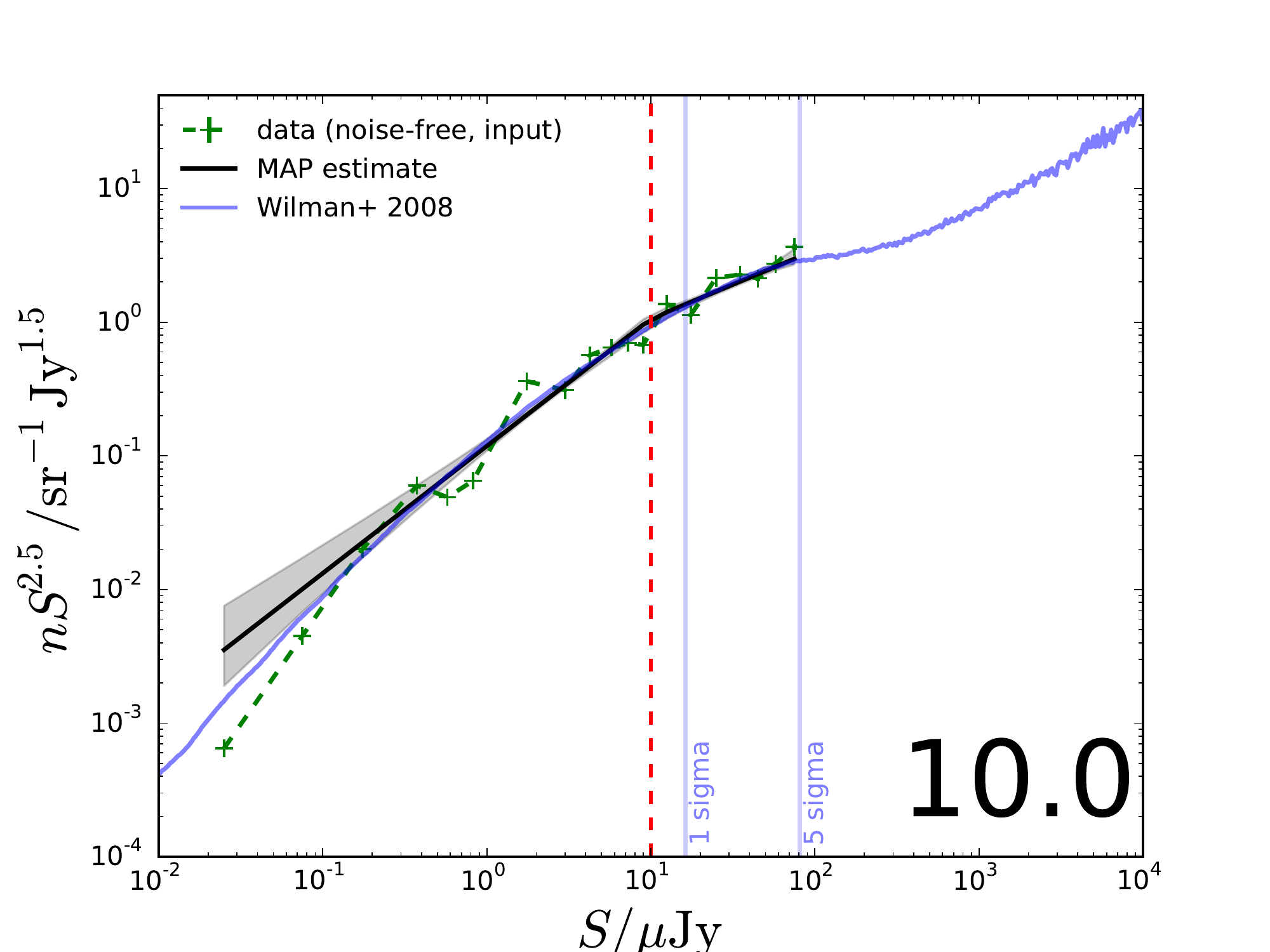}
\includegraphics[width=8.5cm,origin=br,angle=0]{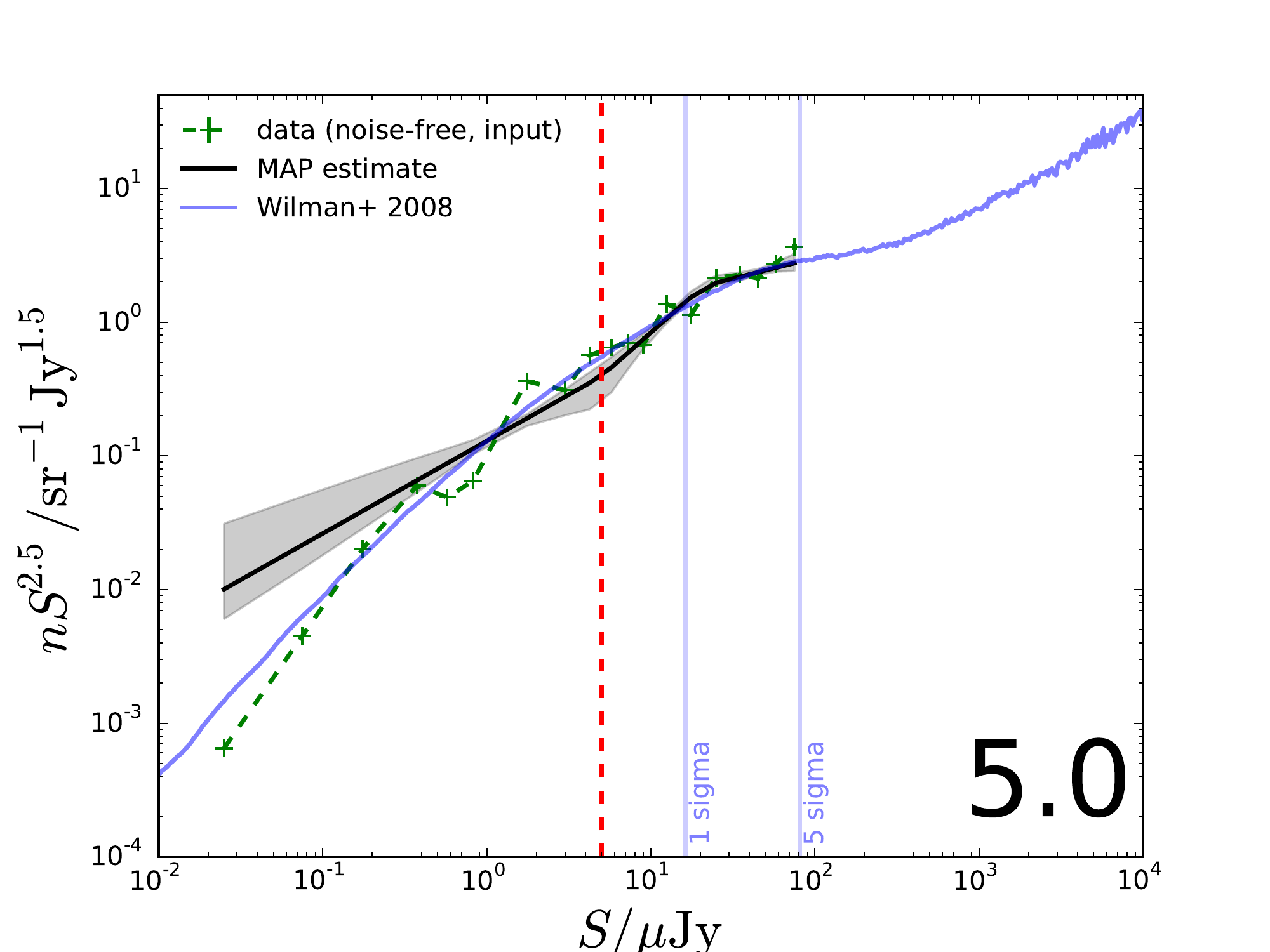}\\
\includegraphics[width=8.5cm,origin=br,angle=0]{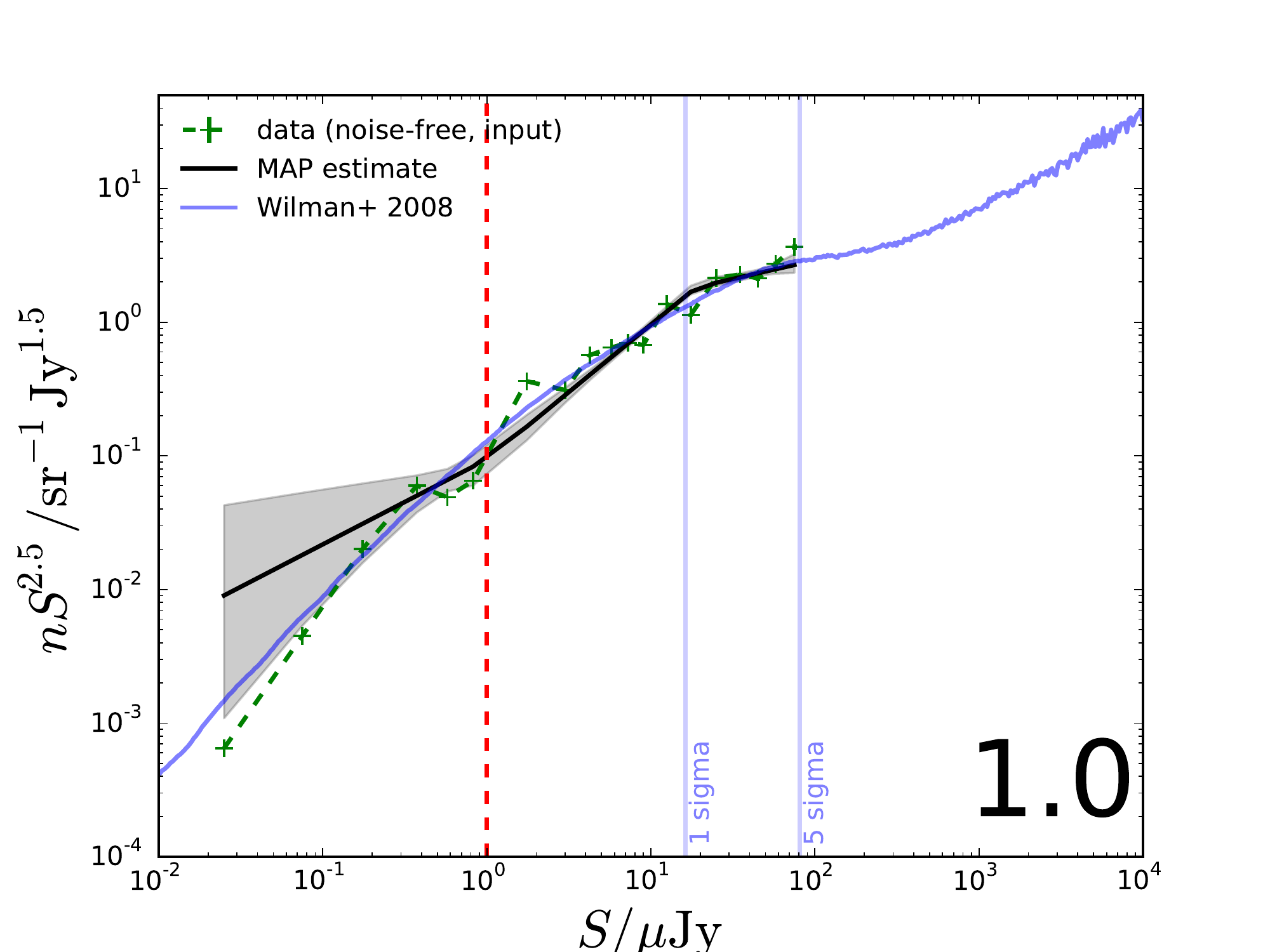}
\includegraphics[width=8.5cm,origin=br,angle=0]{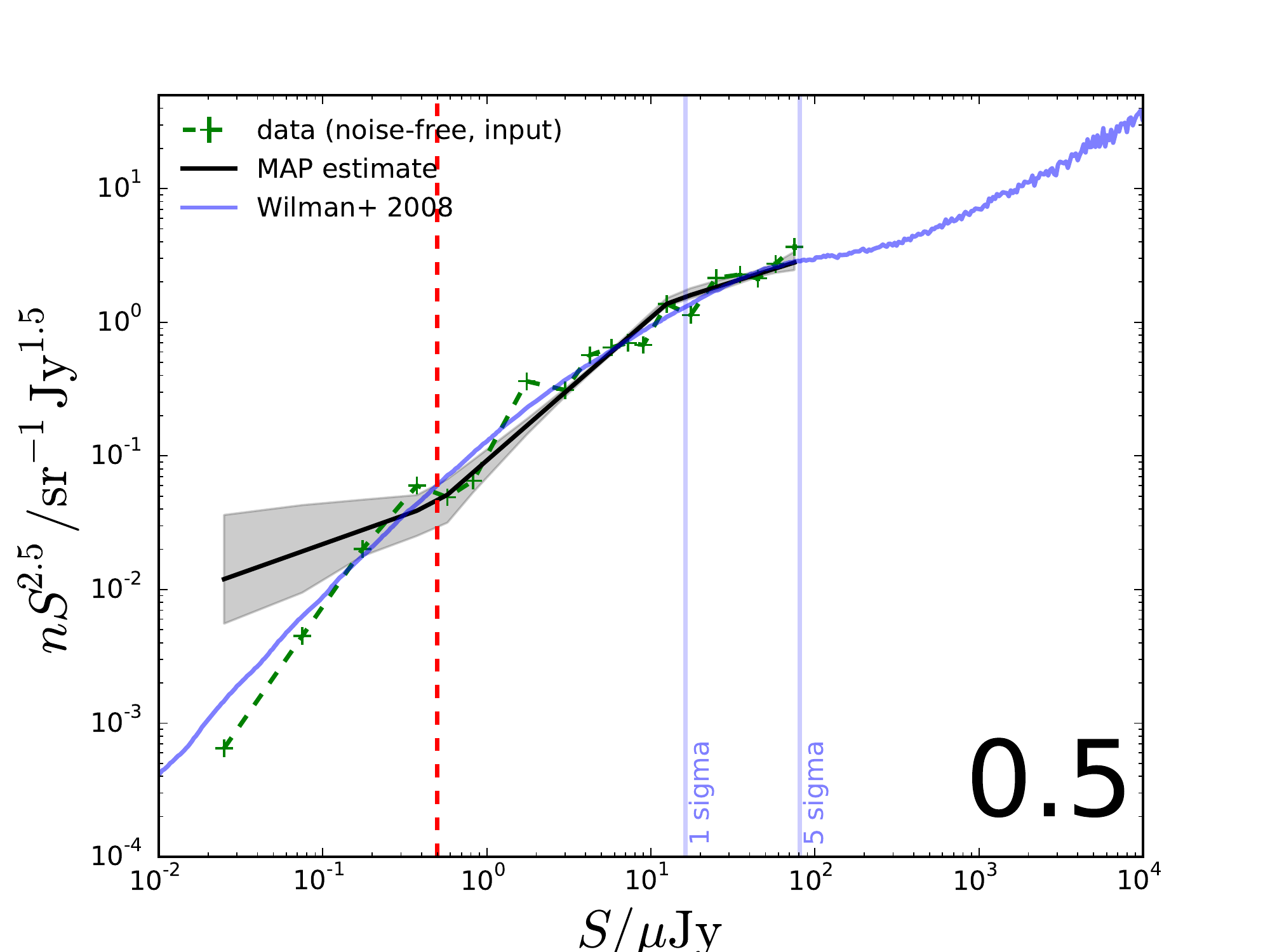}
\caption{For the SKADS catalogue data, source-count reconstruction for
  model C as a function of the flux (indicated by the vertical red
  dashed line) at which the lowest break is fixed: (a) 10, (b) 5, (c) 1 and (d)
  0.5\,$\mu$Jy.
The underlying, noise-free realisation of the counts for
  the 1-deg$^2$ catalogue is indicated by the green dashed line. The real, noisy
  catalogue data are shown as a black dashed line. The black line and
  grey shaded region show the maximum \textit{a posteriori} (MAP) and
  68-per-cent confidence region of the distributions of models reconstructed from
  every sample from the posterior. The
  vertical blue lines are at 1$\sigma$ and 5$\sigma$.
\label{fig:skads-recon2}}
\end{figure*}

To place further constraints on the depth the algorithm can reach, we
fixed the position of the faintest break in model~C and investigated
the effects on the reconstructed counts. Figure~\ref{fig:skads-recon2}
shows what happens as the break is positioned at 10, 5, 1 and
0.5\,$\mu$Jy. The reconstruction is accurate until it breaks down at
$\approx$0.5--1$\mu$Jy, in the sense that the uncertainties suddenly
inflate when the enforced break position is decreased to that flux.
Note that this is consistent with our assertion above that
uncertainties from different bins affect each other, because when the
break position is clamped at a very low flux the uncertainties for the
lowest flux bins reflect the fact that the counts in this
noise-dominated region are independent of the brighter flux bins.

\subsection{Effects of clustering and confusion}
\label{sec:tests:sims:confusion}

Next, in order to account for the effects of clustering and confusion,
we injected all the noise-free SKADS sources into a map of noise,
synthesized beam, size and pixel size equal to those of the VLA map of
section~\ref{sec:obs:radio}. Having injected all the noise-free SKADS
sources into the map, flux extraction proceeds as in
section~\ref{sec:obs:extraction}, for catalogue positions
corresponding to those of injected sources.

\begin{figure*}
\centering
\includegraphics[width=8.5cm,origin=br,angle=0]{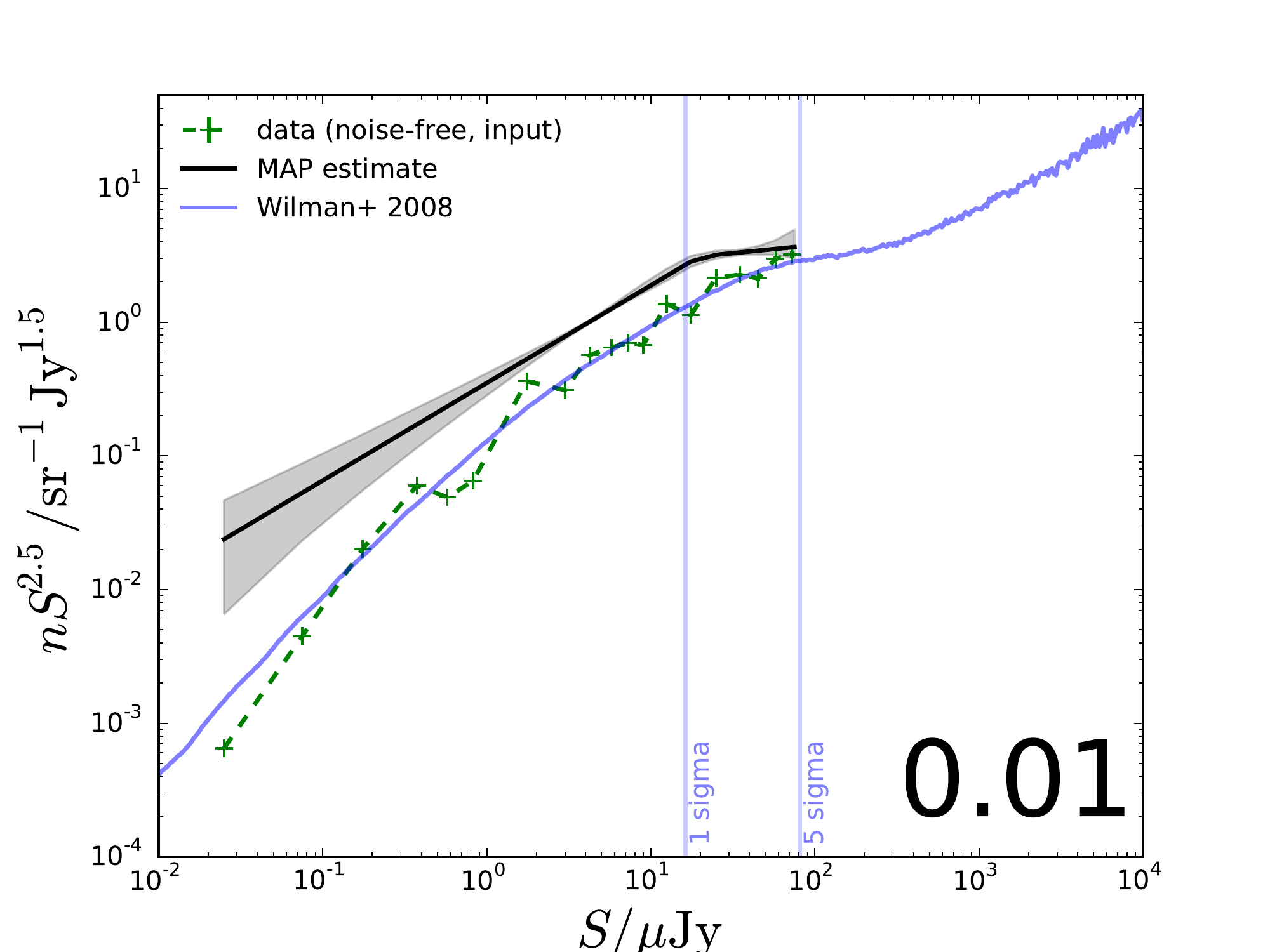}
\includegraphics[width=8.5cm,origin=br,angle=0]{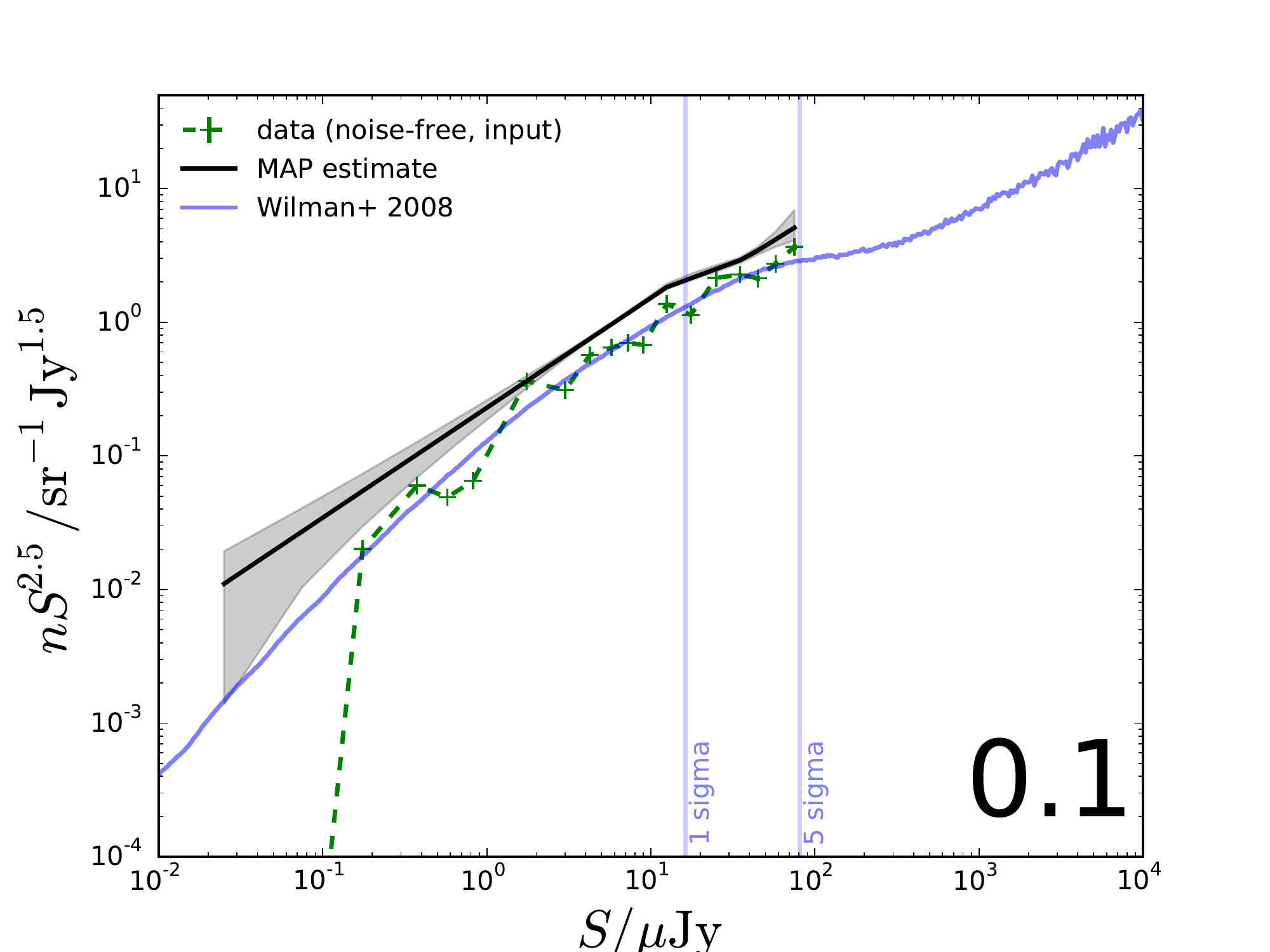}\\
\includegraphics[width=8.5cm,origin=br,angle=0]{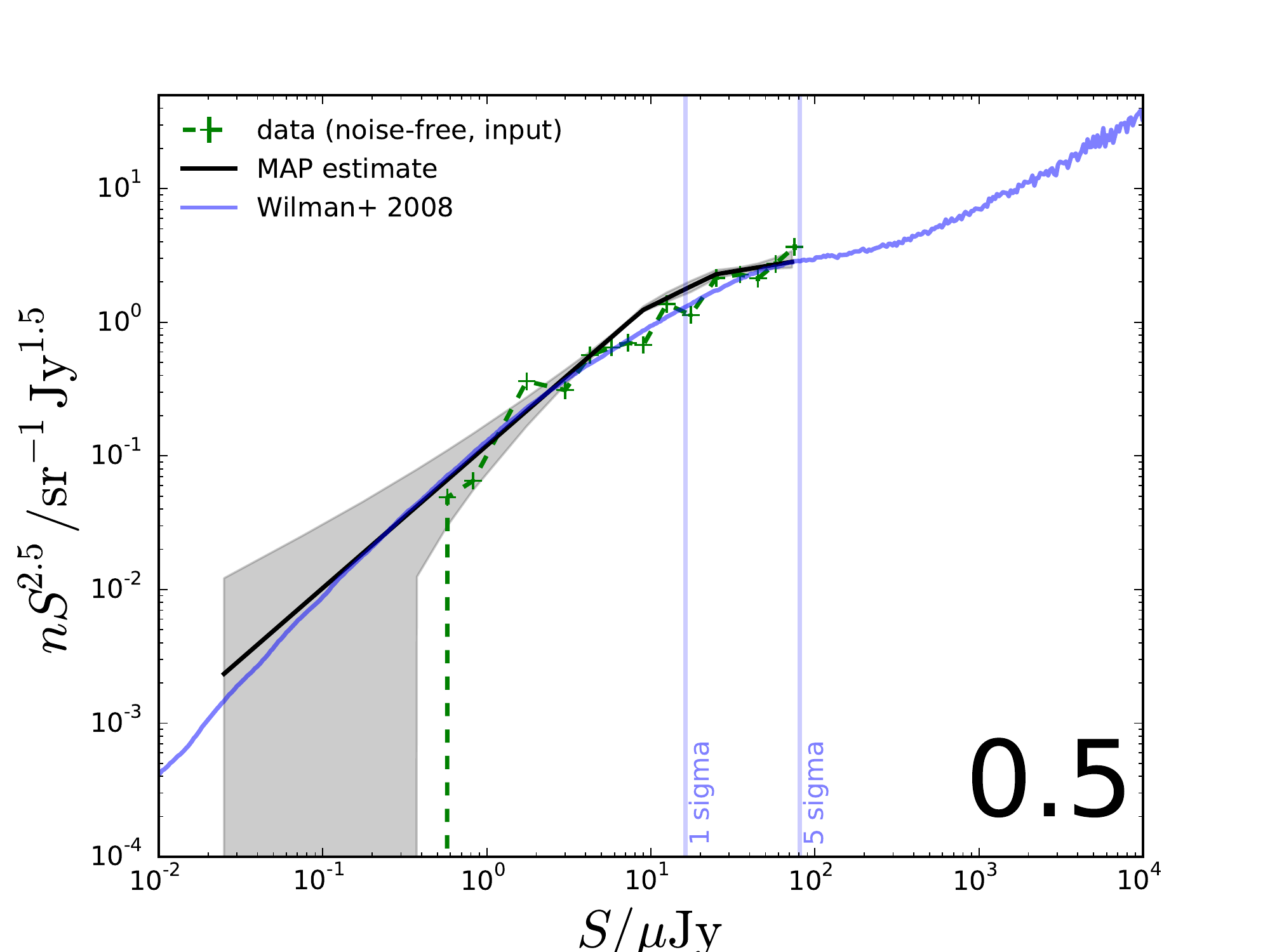}
\includegraphics[width=8.5cm,origin=br,angle=0]{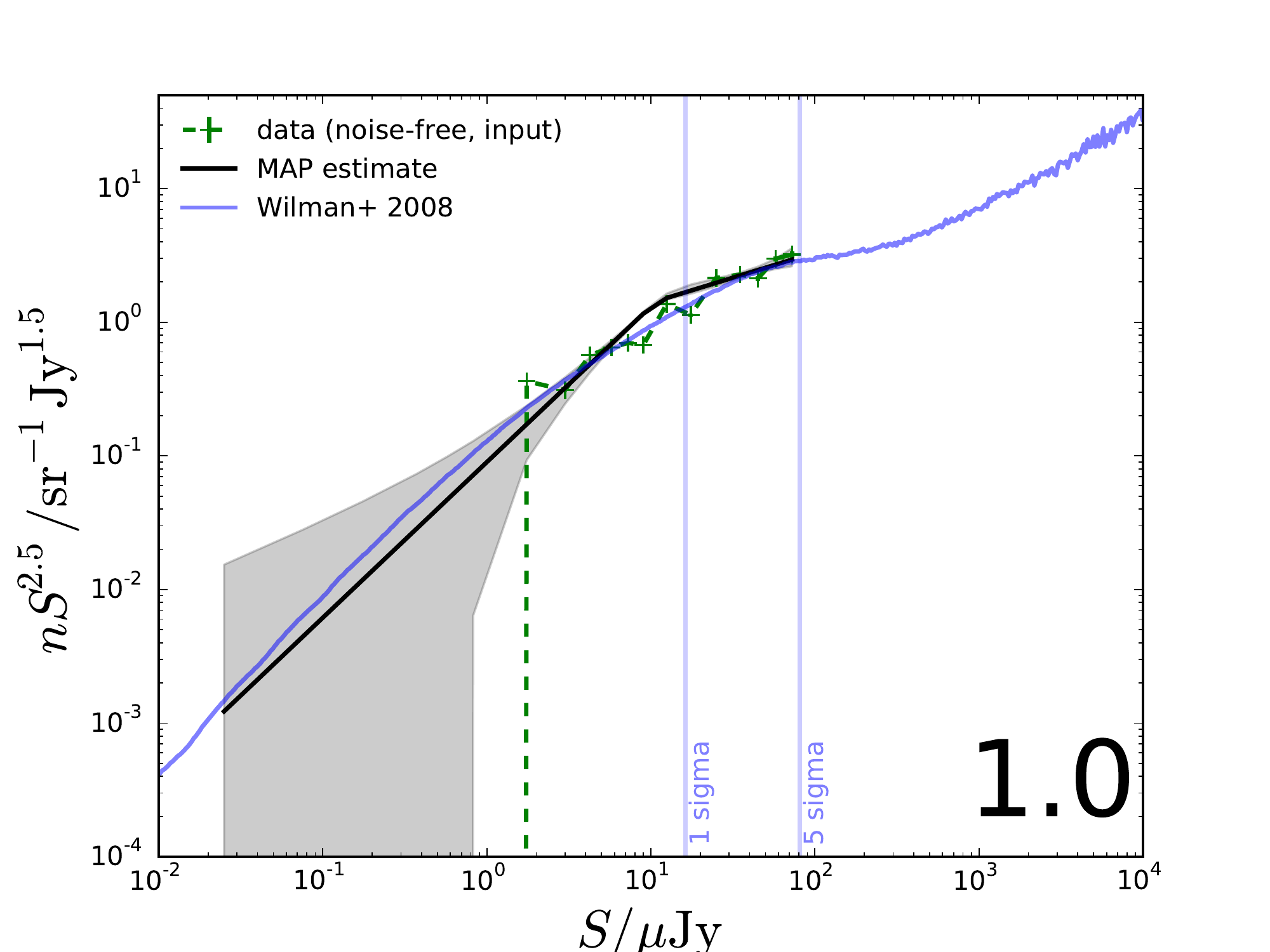}
\caption{Source-count reconstruction for model C as a function of the
  minimum flux injected into the map: (a) 0.01, (b) 0.1, (c) 0.5 and
  (d) 1\,$\mu$Jy.
The underlying, noise-free realisation of the counts for
  the 1-deg$^2$ catalogue is indicated by the green dashed line. The real, noisy
  catalogue data are shown as a black dashed line. The black line and
  grey shaded region show the maximum \textit{a posteriori} (MAP) and
  68-per-cent confidence interval of the distributions of models reconstructed from
  every sample from the posterior. The
  vertical blue lines are at 1$\sigma$ and 5$\sigma$.
\label{fig:skads-recon3}}
\end{figure*}

For this experiment, we fitted model C (with two breaks) to the binned
extracted fluxes. We used model C here because we wanted to allow
sufficient freedom in the possibly now-different reconstruction and it
was only marginally disfavoured over model B. We carried out the fit
with the minimum source flux injected into the map set to be 0.01,
0.1, 0.5 and 1\,$\mu$Jy. Figure \ref{fig:skads-recon3} shows the
reconstructions for these different scenarios. With all sources
$>$10\,nJy injected into the map, the counts are biased high, showing
that, if the SKADS simulations portrayed reality,
confusion would cause the algorithm to break down when applied to
these VLA data even at the 5-$\sigma$ flux limit.
Confusion is still an issue (top right in
Figure~\ref{fig:skads-recon3}) for $S>0.1\mu$Jy, causing the counts to
be artificially boosted, but drops away once all sources $<0.5\mu$Jy
have been excised from the map (bottom left). This is consistent with
the level at which confusion noise is expected in the VLA data (see
section~\ref{sec:obs:radio}).

While potentially disconcerting, this exercise is useful in
understanding the limitations of our modelling scheme. If SKADS is a
true representation of the 1.4-GHz sky, our source counts would be
biased.
If confusion were an issue, efficient deblending algorithms have been
proposed (e.g.~\citealt{kg2010,safarzadeh2015}) to mitigate the
effects of source blending.
We note that this effect may lead to problems for any planned
relatively-low-resolution surveys from SKA pathfinders. We return to
the effects of confusion in
section~\ref{sec:results:float-noise:random}.

\section{Data}
\label{sec:obs}

\subsection{Infrared and Optical Data}
\label{sec:obs:nir}

The VIDEO survey \citep{jarvis2013}, aimed at understanding galaxy
formation, evolution and clusters, is ongoing. The data presented here
cover $\approx$ 1~square degree and are in $Z, Y, J, H,$ and \Ks
bands. The project makes use of Canada--France--Hawaii Telescope
Legacy Survey optical data (CFHTLS--D1; \citealt{ilbert2006}) in the
$u^*, g', r', i', $ and $z'$ bands.

The data selection is similar to that employed by
\cite{zwart2014}.
In order to select galaxies solely by stellar mass (see
\citealt{zwart2014} and discussion therein), we considered just those
with $\Ks<23.5$, i.e.~the 90-per-cent completeness limit for VIDEO's
formal 5-$\sigma$ limit $\Ks=23.8$ (see \citealt{jarvis2013}).  We
further excised any objects contaminated by detector ghosting halos,
as well as any objects that \textsc{SExtractor} deemed to be crowded,
blended, saturated, truncated or otherwise corrupted. 71,418 objects
remained after this data selection, covering an area of
0.97~sq.~deg.

\subsection{Radio Data}
\label{sec:obs:radio}

Radio observations \citep{bondi2003} of the VIDEO field cover a
1-square-degree field centred on J\,2$^h$\,26$^m$\,00$^s$
\,$-$4$\degr$\,30$\arcmin$\,00$\arcsec$ (the XMM--LSS field). The data
comprise a 1.4-GHz Very Large Array (VLA, B-array) 9-point mosaic. For
\textit{any} radio mosaicking strategy, the primary beam unavoidably
imposes a variation in the map noise (see e.g.~\citealt{10C-I}); for
these data this variation is about 20~per~cent around $17.5\,\mu$Jy.
The \textsc{clean} restoring beam is 6\,\arcsec\ full width at half
maximum and the map contains 2048 $\times$ 2048 1.5-arcsec pixels.
In \cite{zwart2014} we estimated the expected confusion noise for
these 6-arcsec-beam data to be $\sigma^*=0.8\,\mu$Jy\,beam$^{-1}$ in
the \cite{condon2012} definition, i.e.~about $\sigma_n/20$ in our
notation.

\subsection{Flux extraction}
\label{sec:obs:extraction}

It was pointed out by \cite{stil2014} that a Nyquist-sampled
synthesized beam could be insufficient for a stacking analysis if the
positional uncertainty exceeds the width of a pixel. In the present
case, the two options were to (i) employ aperture photometry or (ii)
resample the map to a finer grid. Since our near-infra-red field is
relatively crowded, the first would prove difficult without some
deblending scheme (e.g.~\citealt{kg2010,safarzadeh2015}). Hence we
selected the second method, upsampling the VLA map by a factor of 8
using interpolation. The scheme was tested on simulations and on the
detected sources (i.e.~those above $5\sigma$) and the bias in flux
density was able to be reduced to $\lesssim 1$~per~cent.

\begin{figure}
\centering
\includegraphics[width=8cm,origin=br,angle=0]{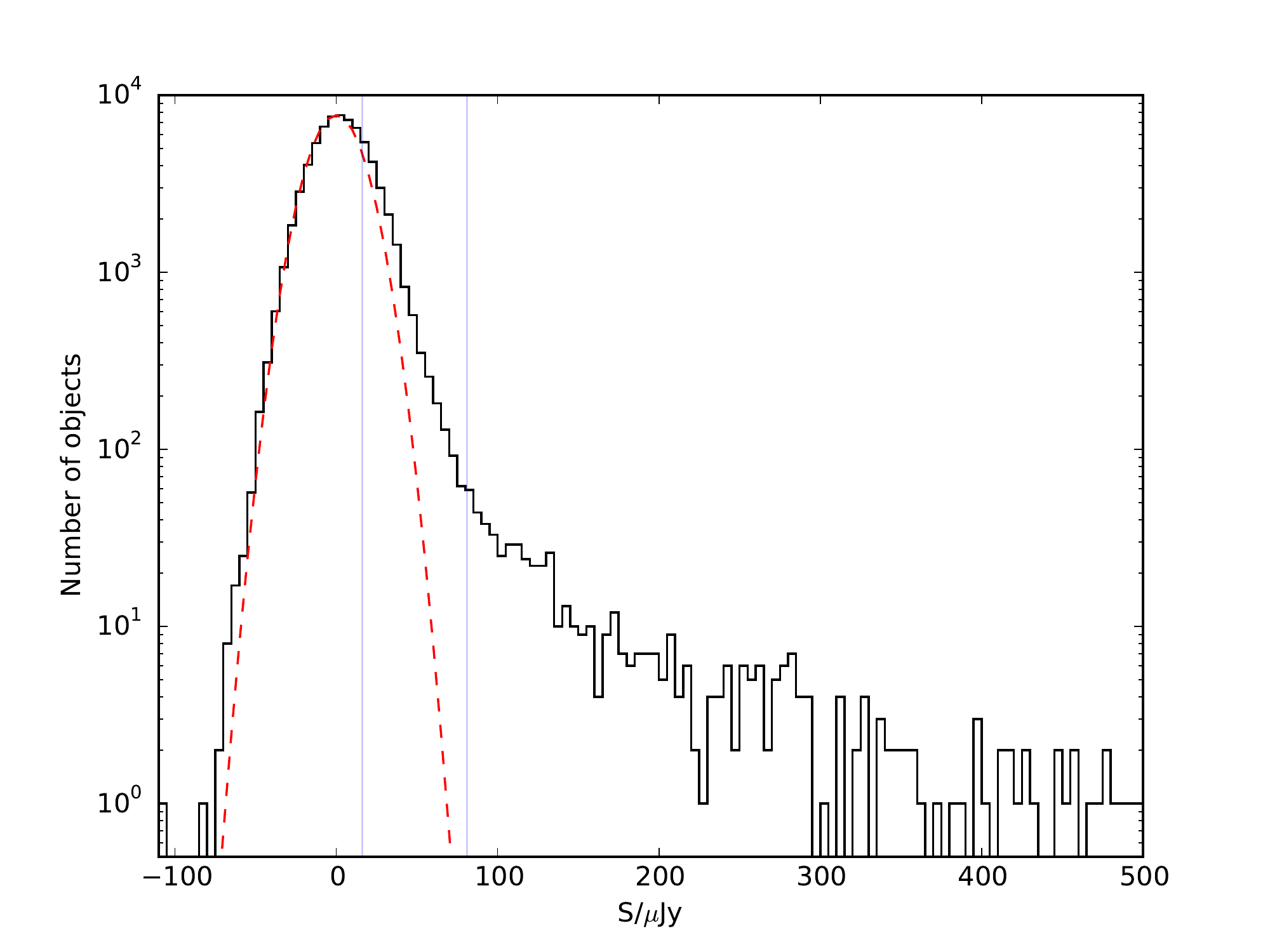}
\caption{Histogram of the fluxes extracted from the VLA map at the
  positions of the VIDEO sources. Overlaid with a red dashed curve is a gaussian of
  width equal to the modal survey noise of 16.2\,$\mu$Jy. The vertical
  blue lines indicate the 1- and 5-$\sigma$ thresholds.\label{fig:flux-histo}}
\end{figure}

A histogram of the extracted fluxes (up to 500\,$\mu$Jy) is shown in
Figure \ref{fig:flux-histo}. Although gaussian on the negative side
(implying domination by thermal noise), there is a long positive tail
caused by discrete radio sources; the tail contains the information
that we will exploit to measure the source-count distribution for
these sources.

\section{Results}
\label{sec:results}

With the algorithm and its implementation fully tested and proven, we
set about extracting binned source counts at the positions of the
VIDEO-detected \Ks-selected sources as described in
section~\ref{sec:obs:extraction}. We used 41 bins in the range
$-108<S/\mu$Jy $<85$, with tighter binning near zero to reflect the
increased numbers of sources in that flux region.

The relative log$_{\mathrm{e}}$-evidences (for fixed $\sigma_n$) are given in
Table~\ref{table:data-evs}. The preferred model, i.e.~the one with the
highest evidence, is model~D.
The posterior probability distribution for that model is given in
Figure~\ref{fig:video-tri}. Figure \ref{fig:recon-results-all} shows
the reconstructions for the four models. A single power law (model A)
has a slope that is consistent with that from \cite{wilman2008} and
\cite{vernstrom2014}, but the amplitude is relatively low and the fit
at the bright end is poor. The evidence anyway strongly rejected this
model. When a break is added (model B), the fit at the bright end
becomes consistent with the $>5\sigma$ counts from \cite{bondi2003}.

The model~C and~D reconstructions are very similar, and our
conclusions remain the same even though D is formally preferred.
Taking model~C (with two breaks), then, the counts remain flat to just
below 20$\mu$Jy ($\simeq 1\sigma$), before falling steeply. After the
second break, at around 3$\mu$Jy, the counts become shallower and do
not become consistent with those from \cite{wilman2008} and
\cite{vernstrom2014} again until $<0.5\mu$Jy.

\subsection{Sensitivity to thermal-noise contribution}
\label{sec:results:float-noise}

As a final check, we repeated the same fitting procedure, but allowing
for an extra degree of freedom in that $\sigma_n$ was now allowed to
vary uniformly in the range
$0.5$--$2.0\times\sigma_{\mathrm{survey,modal}}$. This was because
there was a hint (see for example Figure~\ref{fig:flux-histo}) that
the assumed map noise was an underestimate.
Table~\ref{table:data-evs2} shows the model-selection results.
Immediately one sees that the relative evidences are overwhelmingly
higher for this family of models when compared to the fixed-noise set.
Model~B\arcmin\ is now the preferred model, though only marginally
compared to A\arcmin, C\arcmin\ and D\arcmin, but it is nonetheless
model~B\arcmin\ that we formally adopt.
Figure~\ref{fig:recon-results-all2} gives the corresponding
reconstructions and Figure~\ref{fig:video-tri2} is the posterior
distribution for model~B\arcmin. The intermediate break in the counts
now disappears, indicating that the earlier models had probably been
compensating for an incorrect noise figure, the model now being
summarizable as flat to $\approx 40\mu$Jy, then falling with a slope
of $-$1.7 below that flux. The map noise estimated via the new route
is $17.5\pm 0.08\mu$Jy rather than the modal value of 16.2$\mu$Jy
assumed earlier.

\subsection{Interpretation}
\label{sec:results:literature}

\cite{hjc2013} showed that the scatter in the counts at 100$\mu$Jy
could be attributed to sample variance, but this becomes less true at
fainter fluxes as the raw numbers of sources increase, so that it is
unclear that this could scatter our counts. Assuming, then, no
contribution from sample variance,
we might interpret the very slight deficit of radio emission in that
region to indicate that some small fraction of the
\cite{vernstrom2014} radio sources are beyond the faintness limit
of the VIDEO data. Note that the difference in the counts cannot be
due to confusion, since we found earlier that confusion tended to bias
the counts \textit{upwards}. We see also that our inferred source
counts are not consistent with those derived from the ARCADE2 data
(\citealt{arcade2} and \citealt{arcade2-2}), which afforded integral
constraints on the 3--90-GHz temperature contribution of a putative
population of non-Galactic discrete radio sources below some limiting
flux, constraining the differential counts to be $\propto S^{-2.6}$,
i.e.~very flat in the Euclidean normalization. Our results thus give
further weight to the argument of \cite{vernstrom2014} (see their
Figure~18) that the ARCADE2 `excess' cannot be accounted for by a
population of faint discrete sources.

\subsection{Could there be a confusing background?}
\label{sec:results:float-noise:random}

As a final test, we undertook map extractions at 72,000
\textit{random} positions for both the SKADS simulations and the
VIDEO-VLA data. This should give an indication of the confusion
`background' away from the stacked positions.
No $>5\sigma$ sources were injected into the SKADS map. In the case of
the VIDEO data, we masked out the known (i.e.~$>5\sigma$) sources from
the \cite{bondi2003} catalogue using circular apertures of radius
$5\times$ the full width at half maximum of the synthesized beam,
random pixels being drawn without replacement from the remaining area
(subsequently corrected for).

\begin{figure*}
\centering
  \includegraphics[width=8.5cm,origin=br,angle=0]{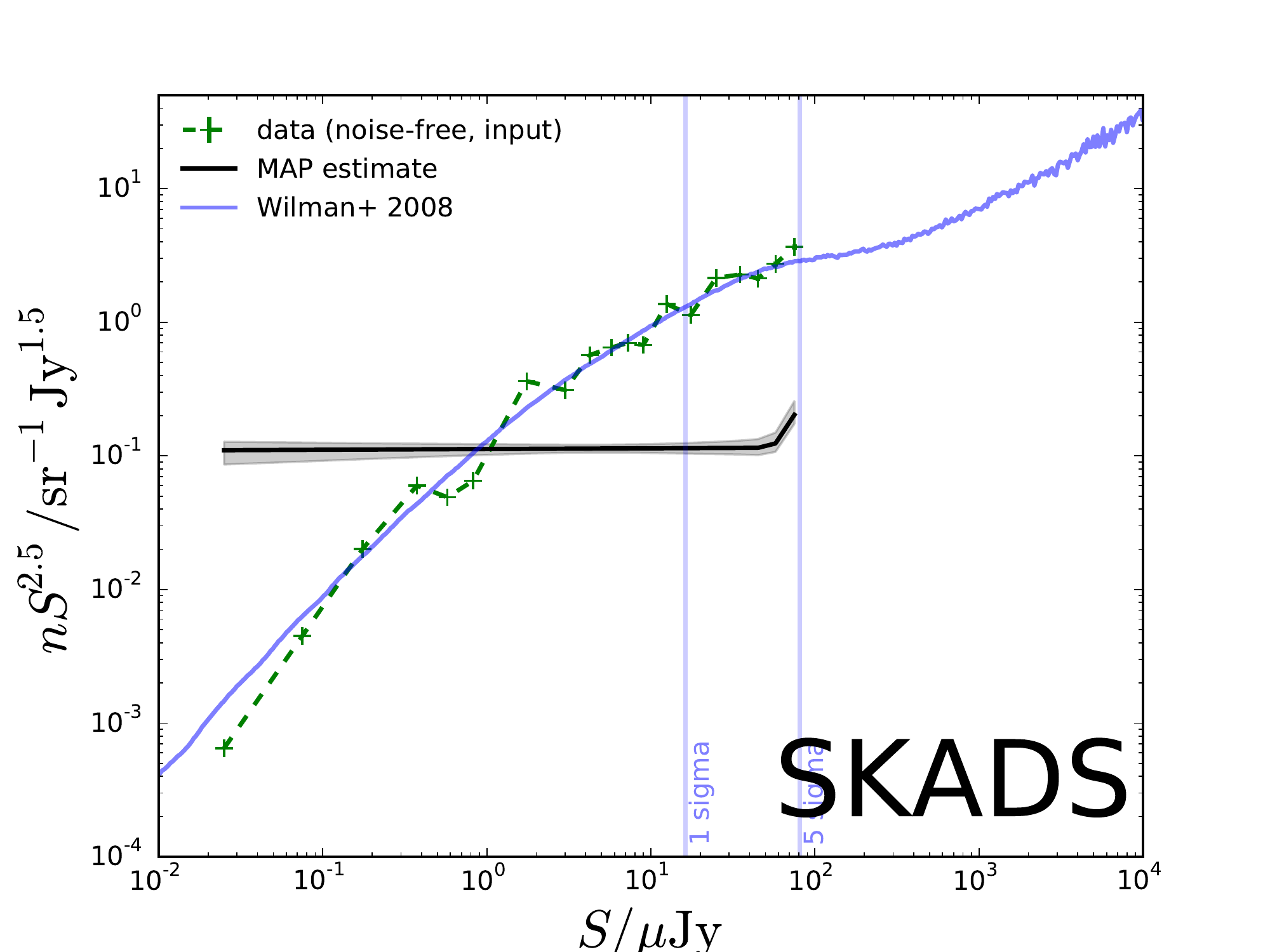}
\includegraphics[width=8.5cm,origin=br,angle=0]{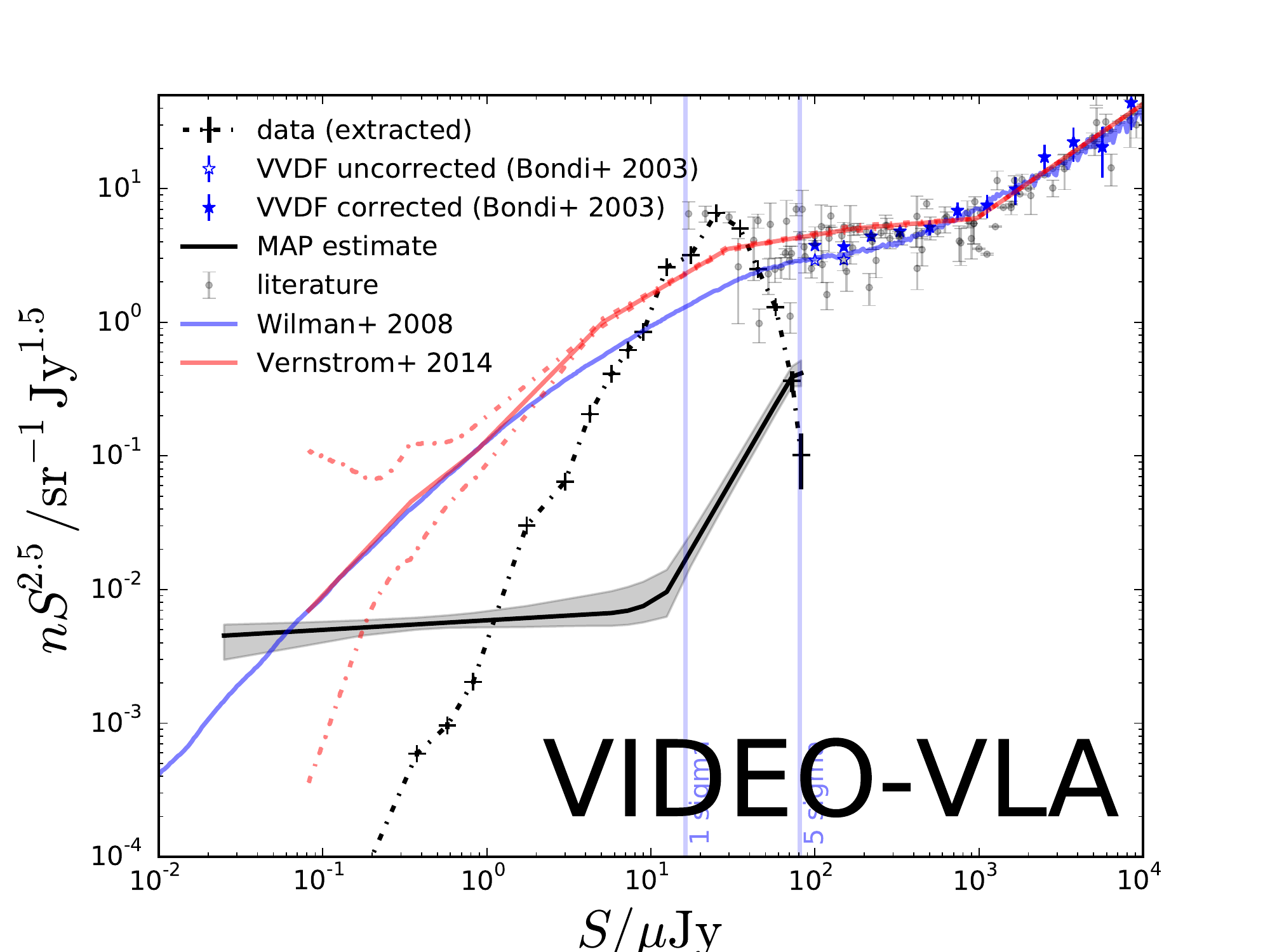}
  \caption[]{Source-count reconstructions for the preferred models for
    SKADS (left, model~B) and VIDEO (right, model~D\arcmin) when
    72,000 random map positions are used for the stacking.
The underlying, noise-free realisation of the counts for
  the 1-deg$^2$ catalogue is indicated by the green dashed line. The real, noisy
  catalogue data are shown as a black dashed line. The black line and
  grey shaded region show the maximum \textit{a posteriori} (MAP) and
  68-per-cent confidence interval of the distributions of models reconstructed from
  every sample from the posterior. The
  vertical blue lines are at 1$\sigma$ and 5$\sigma$.
 The underlying full SKADS
  \citep{wilman2008} counts are shown in blue, and the only other
  counts at these depths \citep{vernstrom2014} are marked in red, with
  the 68~per~cent confidence interval in dashed red. The filled and
  empty blue stars respectively represent the \cite{bondi2003}
  corrected and uncorrected counts for the VLA $>5\sigma$ sources.
  Literature values\protect\footnotemark\ are taken from the review by
  \cite{dezotti-review}.\label{fig:recon-results-random}}
\end{figure*}

\begin{figure*}
\centering
\includegraphics[width=1.0\textwidth,origin=br,angle=0]{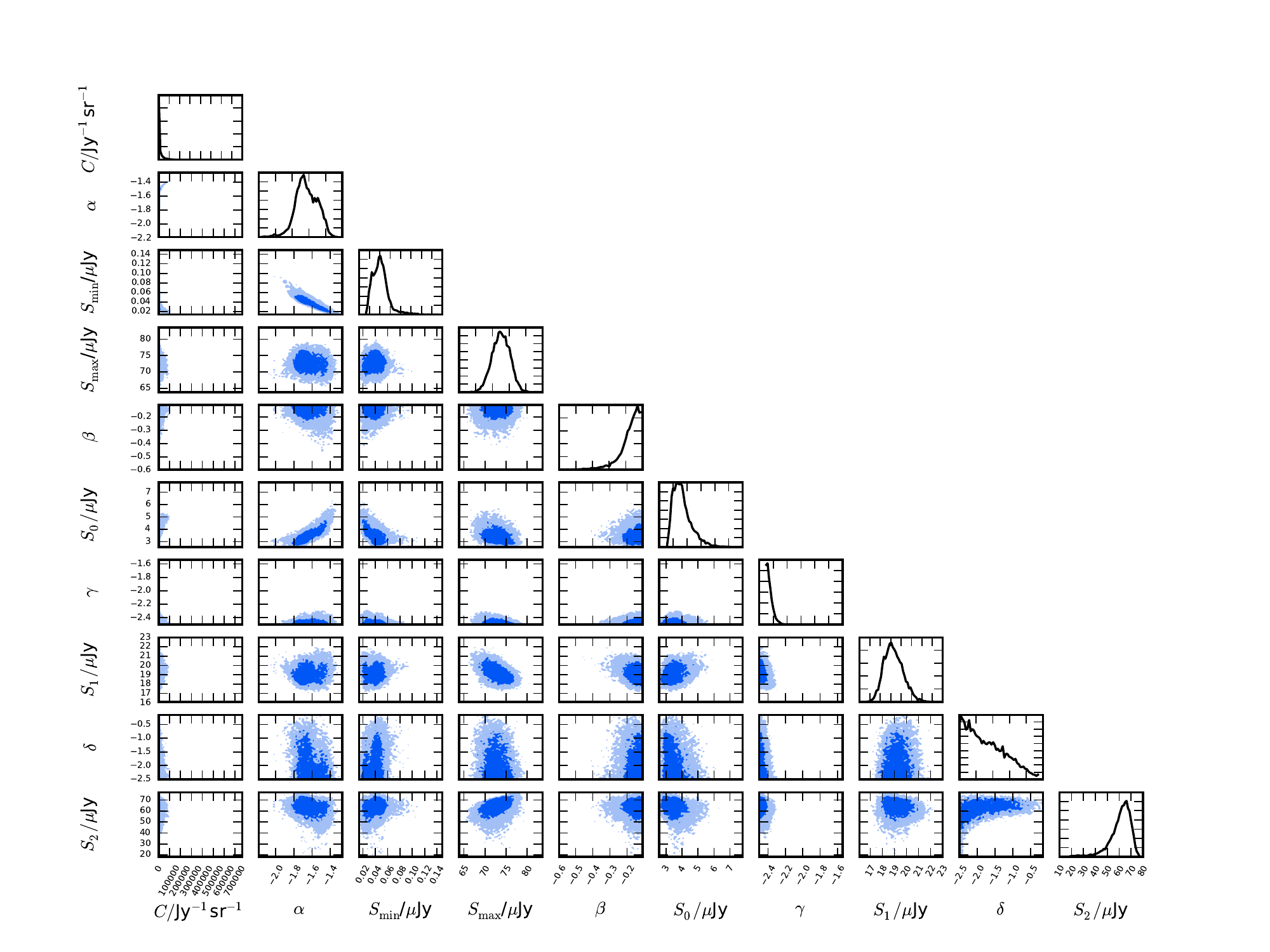}
\caption{Posterior probability distribution for the three-break
  model~D fitted to the VIDEO data. The dark and light shaded regions
  indicate the 68 and 95~per~cent confidence
  limits.\label{fig:video-tri}}
\end{figure*}

\begin{figure*}
\centering
\includegraphics[width=8.5cm,origin=br,angle=0]{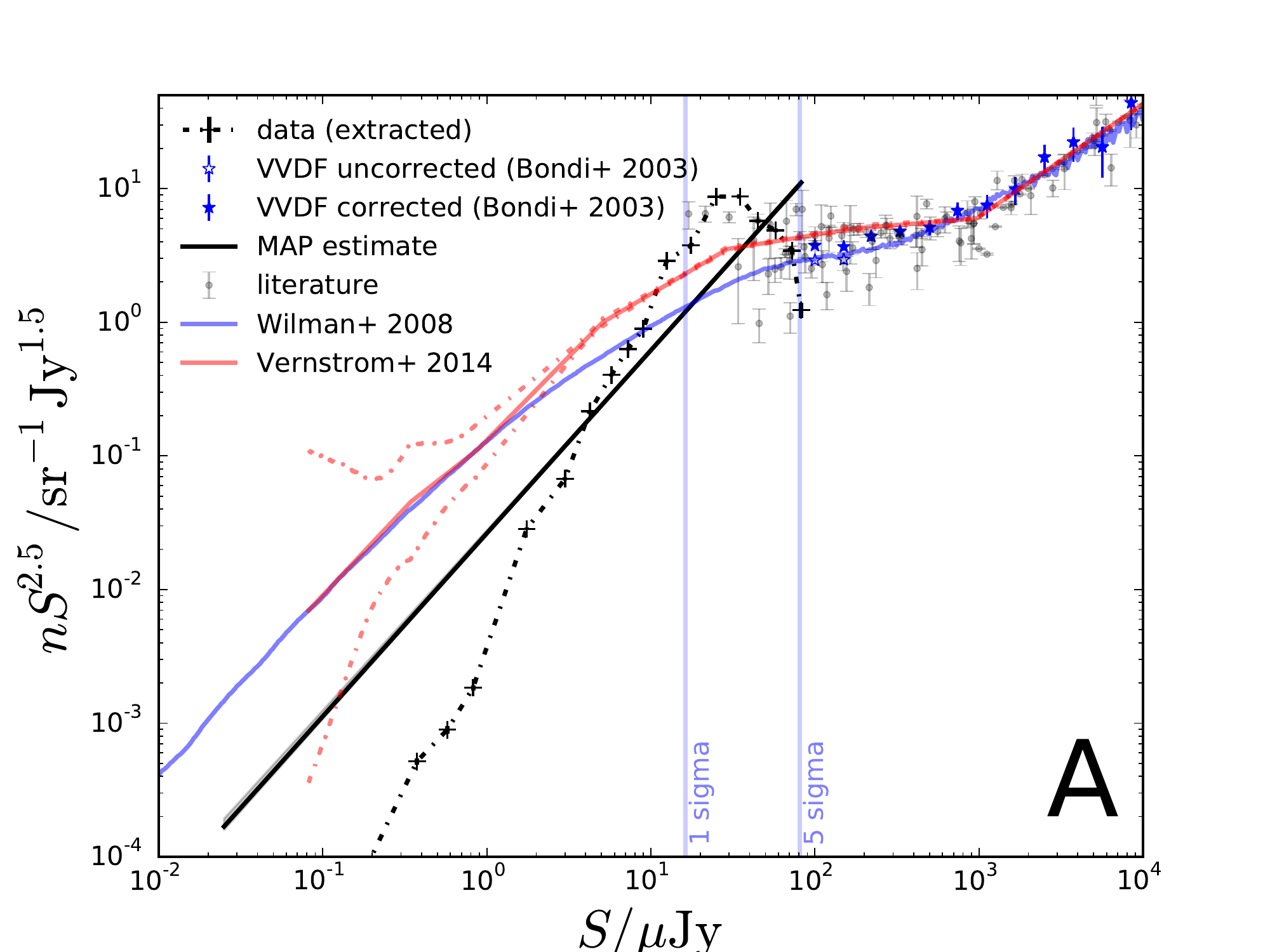}
\includegraphics[width=8.5cm,origin=br,angle=0]{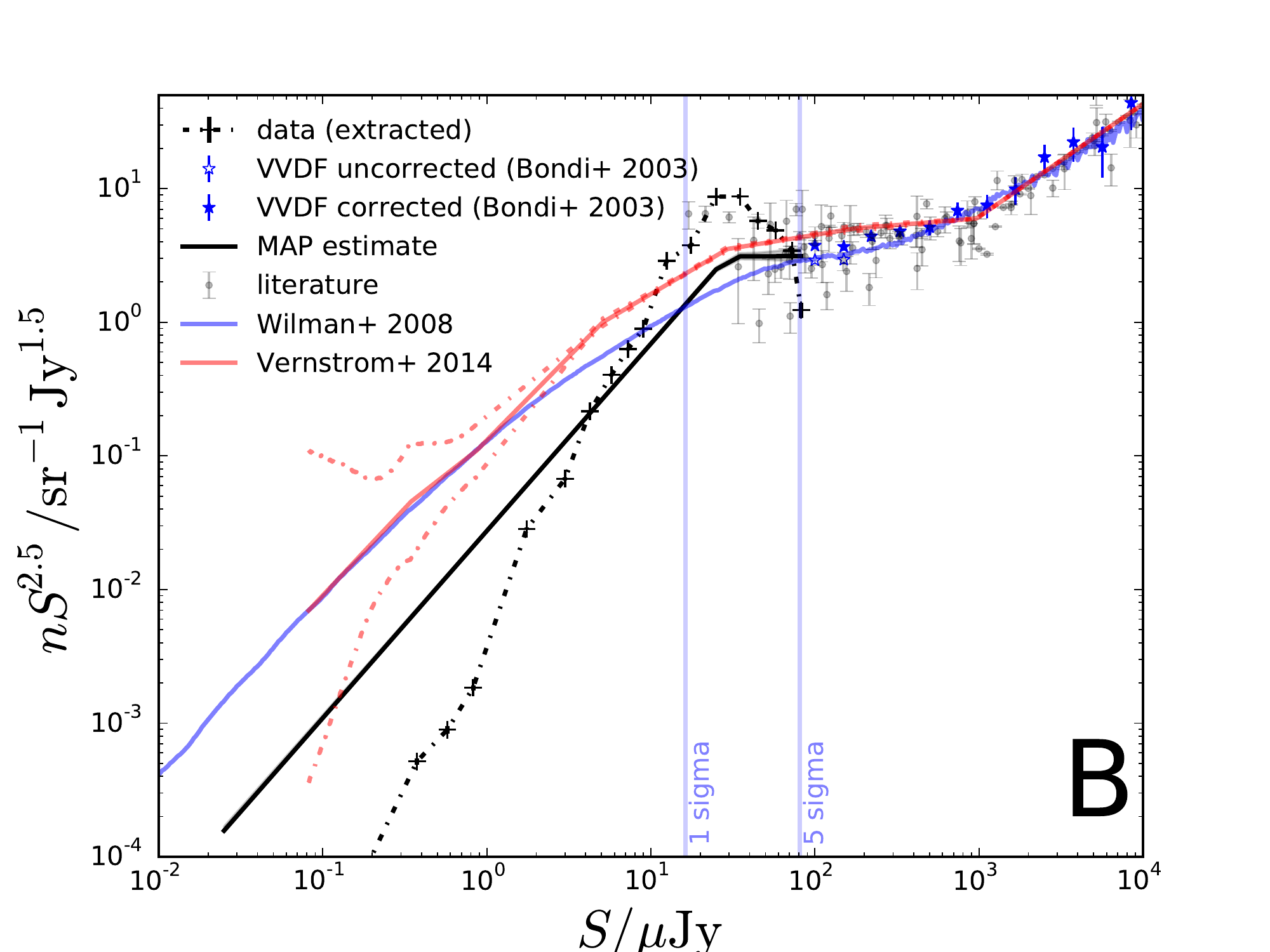}\\
\includegraphics[width=8.5cm,origin=br,angle=0]{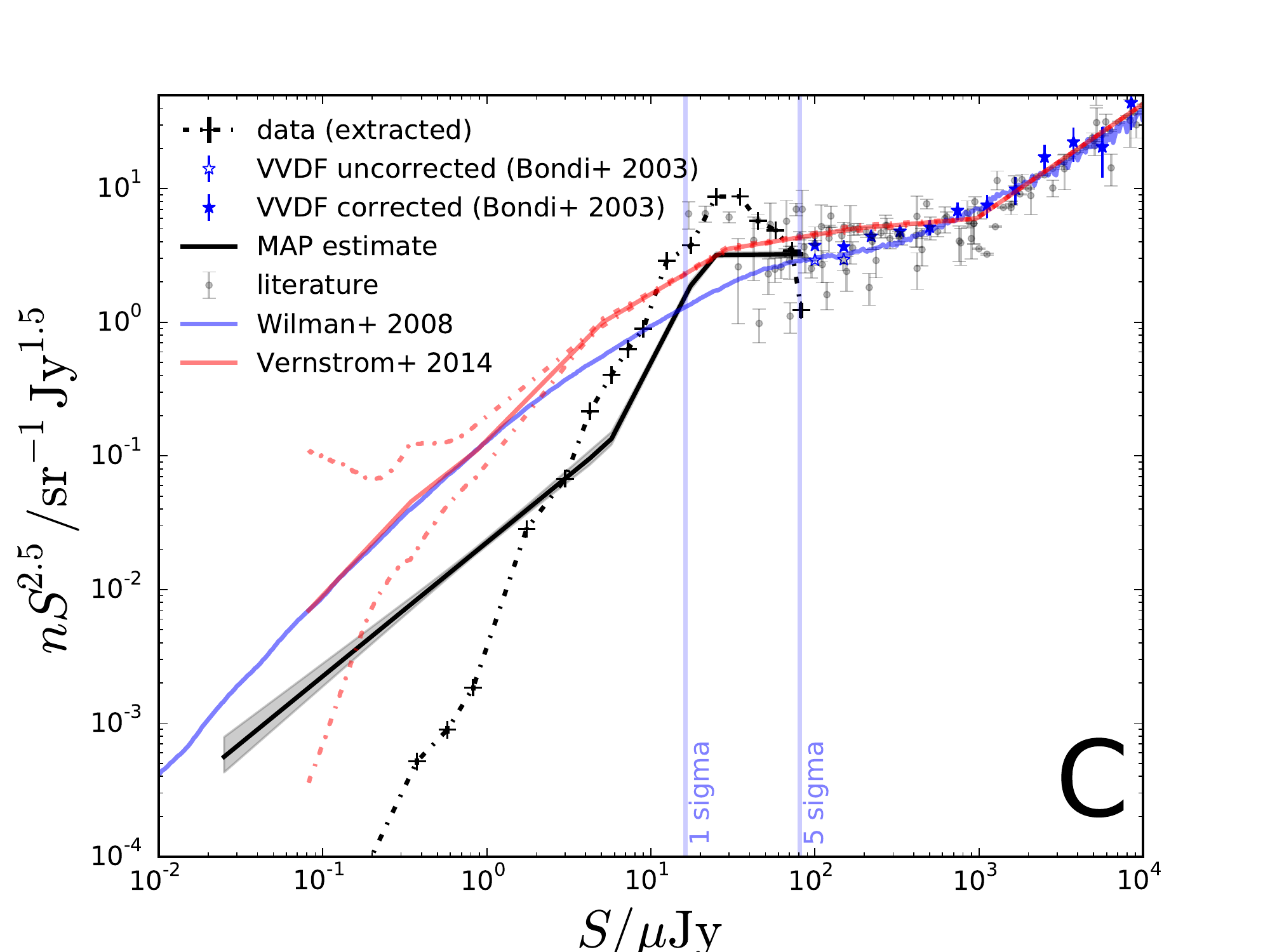}
\includegraphics[width=8.5cm,origin=br,angle=0]{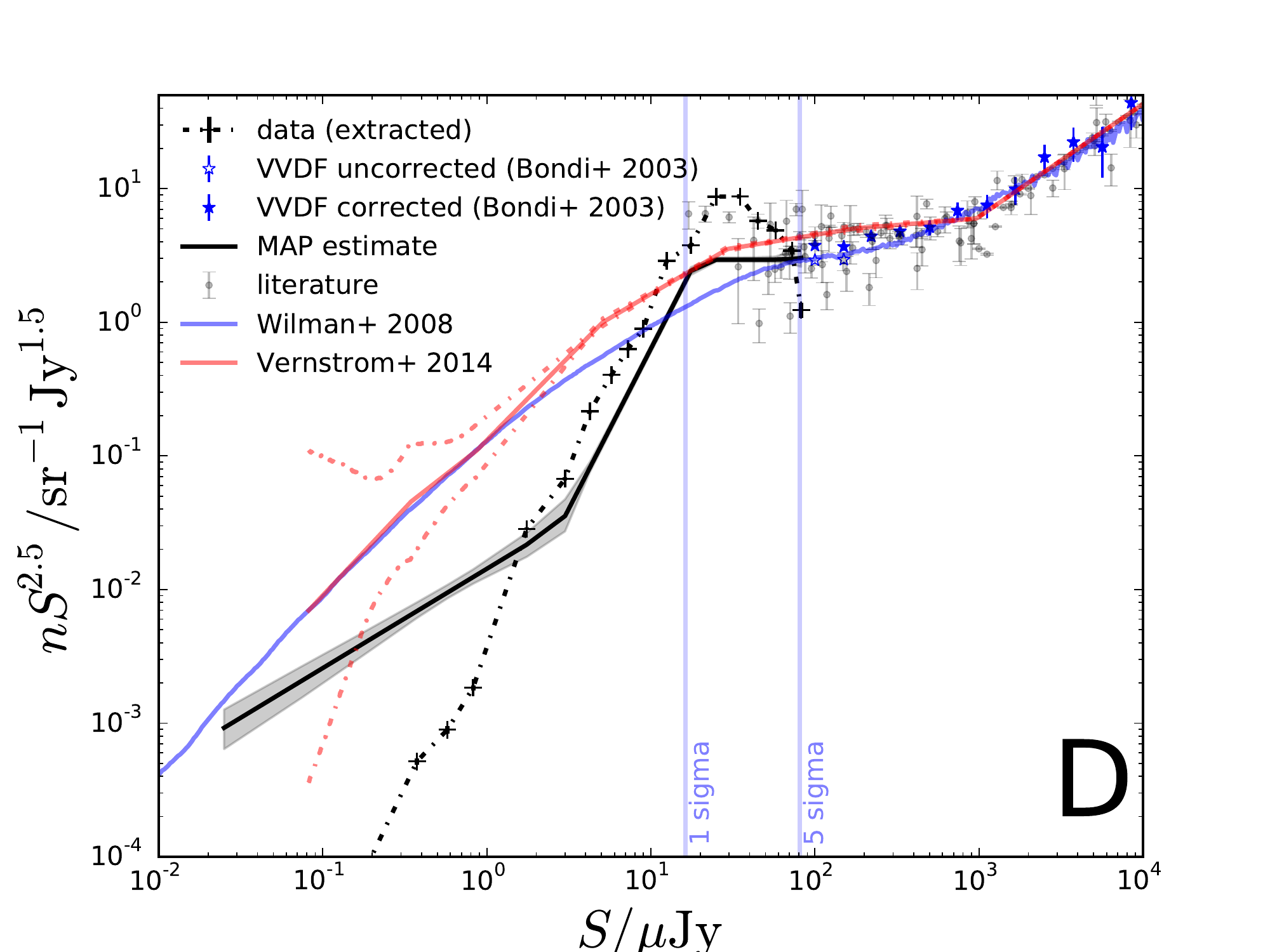}
\caption[]{Reconstructions for all galaxies in the VIDEO sample. The
  model fitted is indicated bottom right. The underlying full SKADS
  \citep{wilman2008} counts are shown in blue, and the only other
  counts at these depths \citep{vernstrom2014} are marked in red, with
  the 68~per~cent confidence interval in dashed red. The filled and
  empty blue stars respectively represent the \cite{bondi2003}
  corrected and uncorrected counts for the VLA $>5\sigma$ sources.
  Literature values are taken from the review by
  \cite{dezotti-review}.
The real, noisy data are shown as a black dashed line. The black line and
  grey shaded region show the maximum \textit{a posteriori} (MAP) and
  68-per-cent confidence interval of the distributions of models reconstructed from
  every sample from the posterior. The
  vertical blue lines are at 1$\sigma$ and 5$\sigma$.\label{fig:recon-results-all}}
\end{figure*}

\begin{figure*}
\centering
\includegraphics[width=8.5cm,origin=br,angle=0]{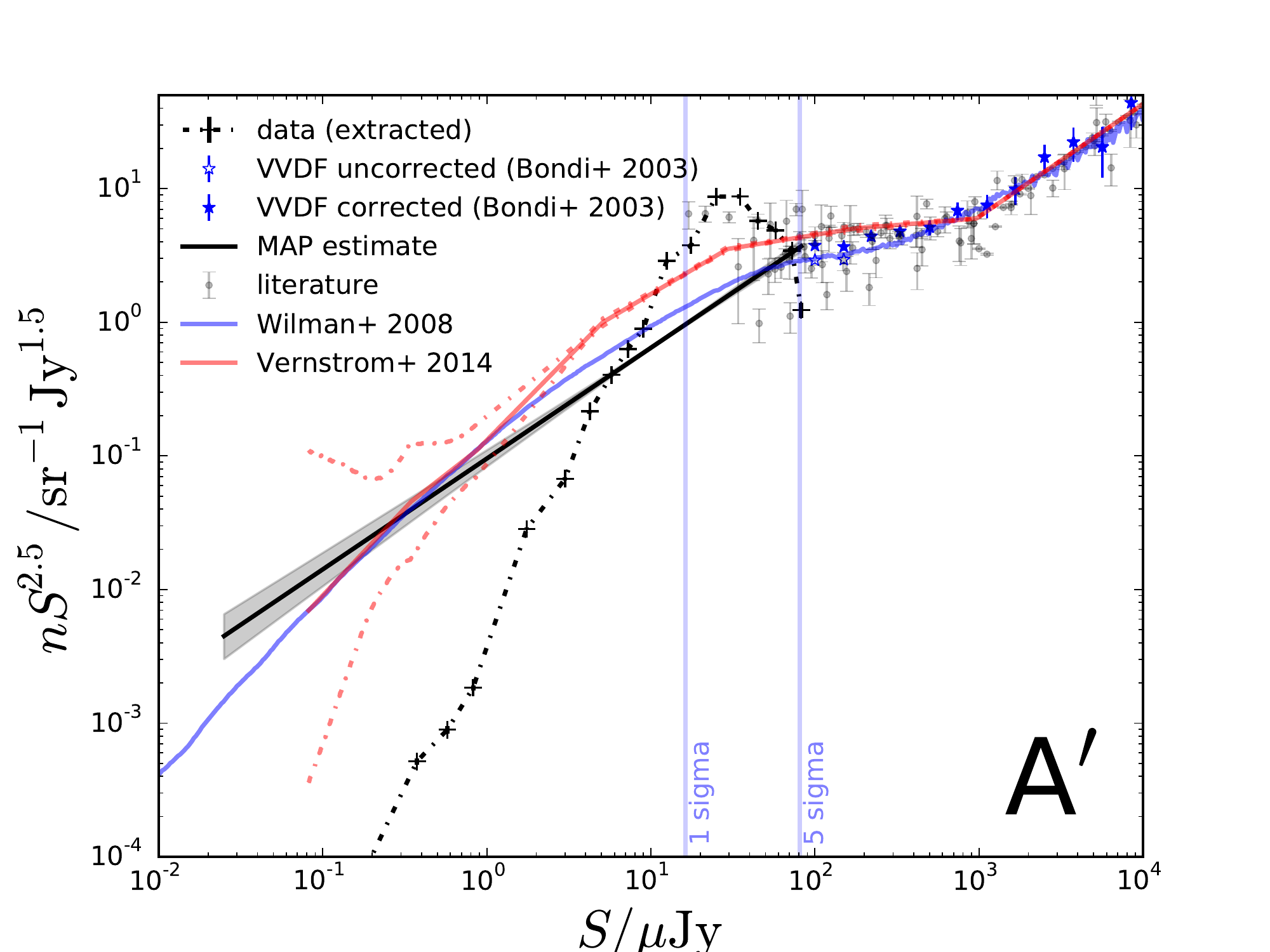}
\includegraphics[width=8.5cm,origin=br,angle=0]{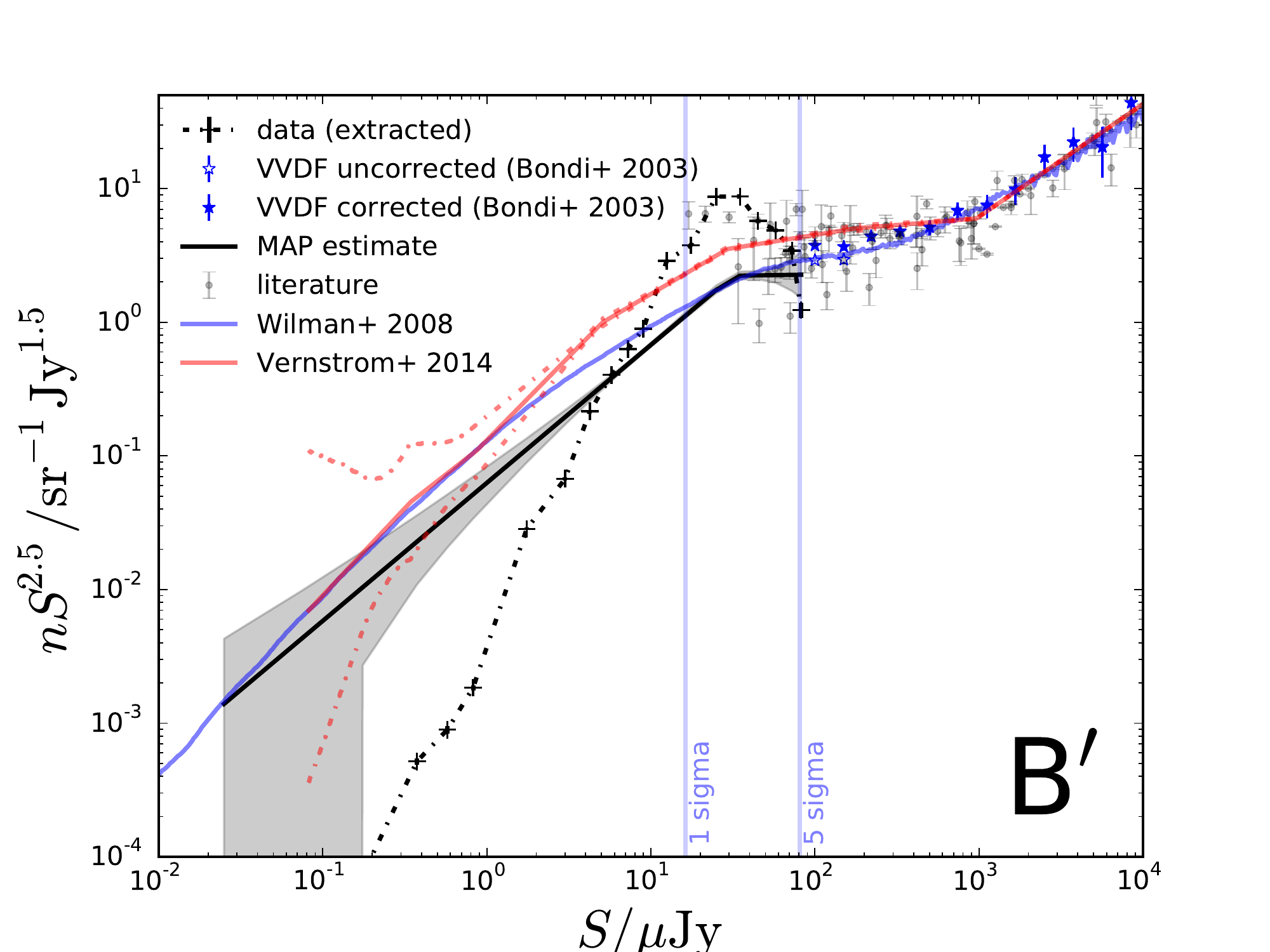}\\
\includegraphics[width=8.5cm,origin=br,angle=0]{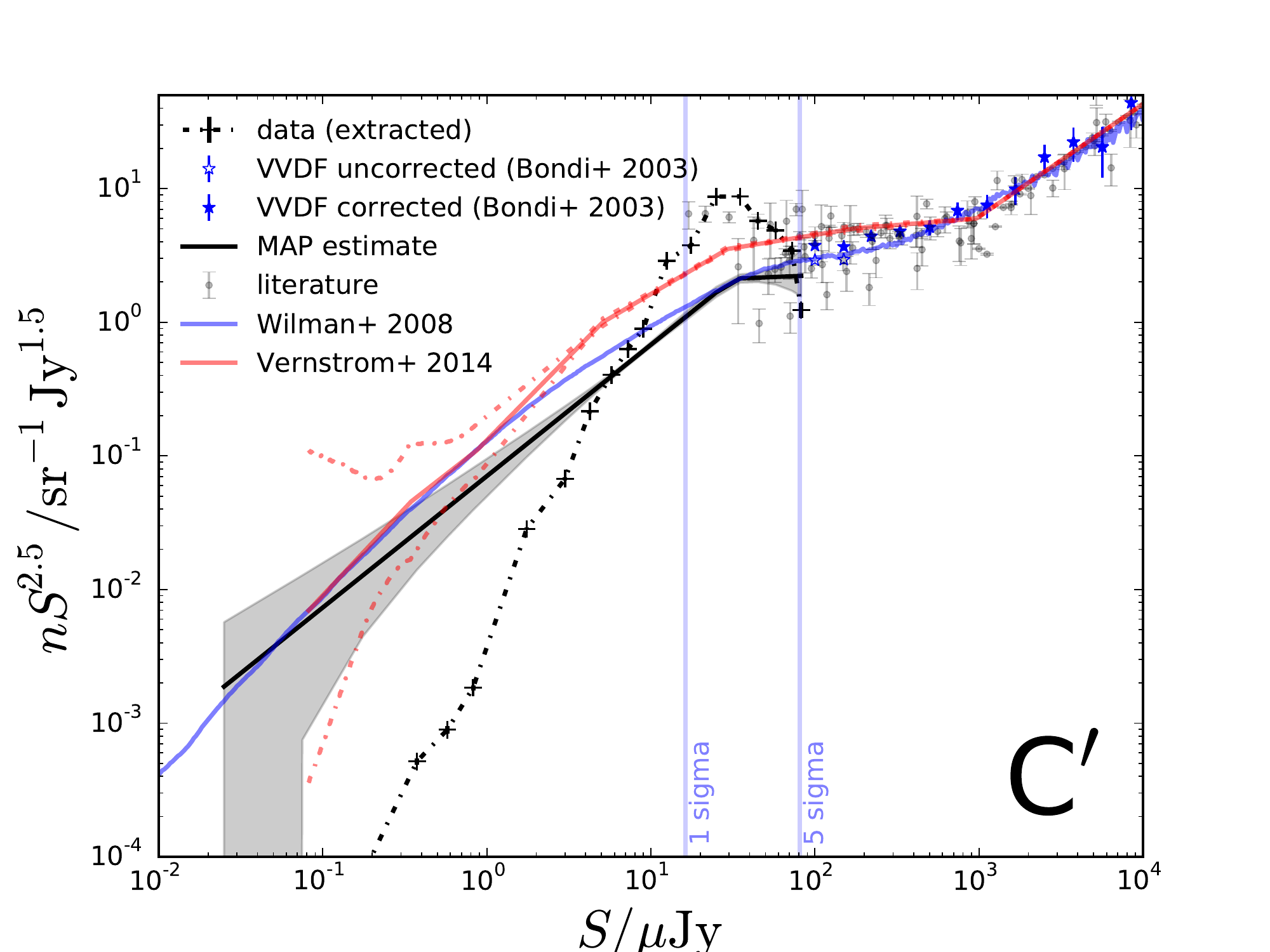}
\includegraphics[width=8.5cm,origin=br,angle=0]{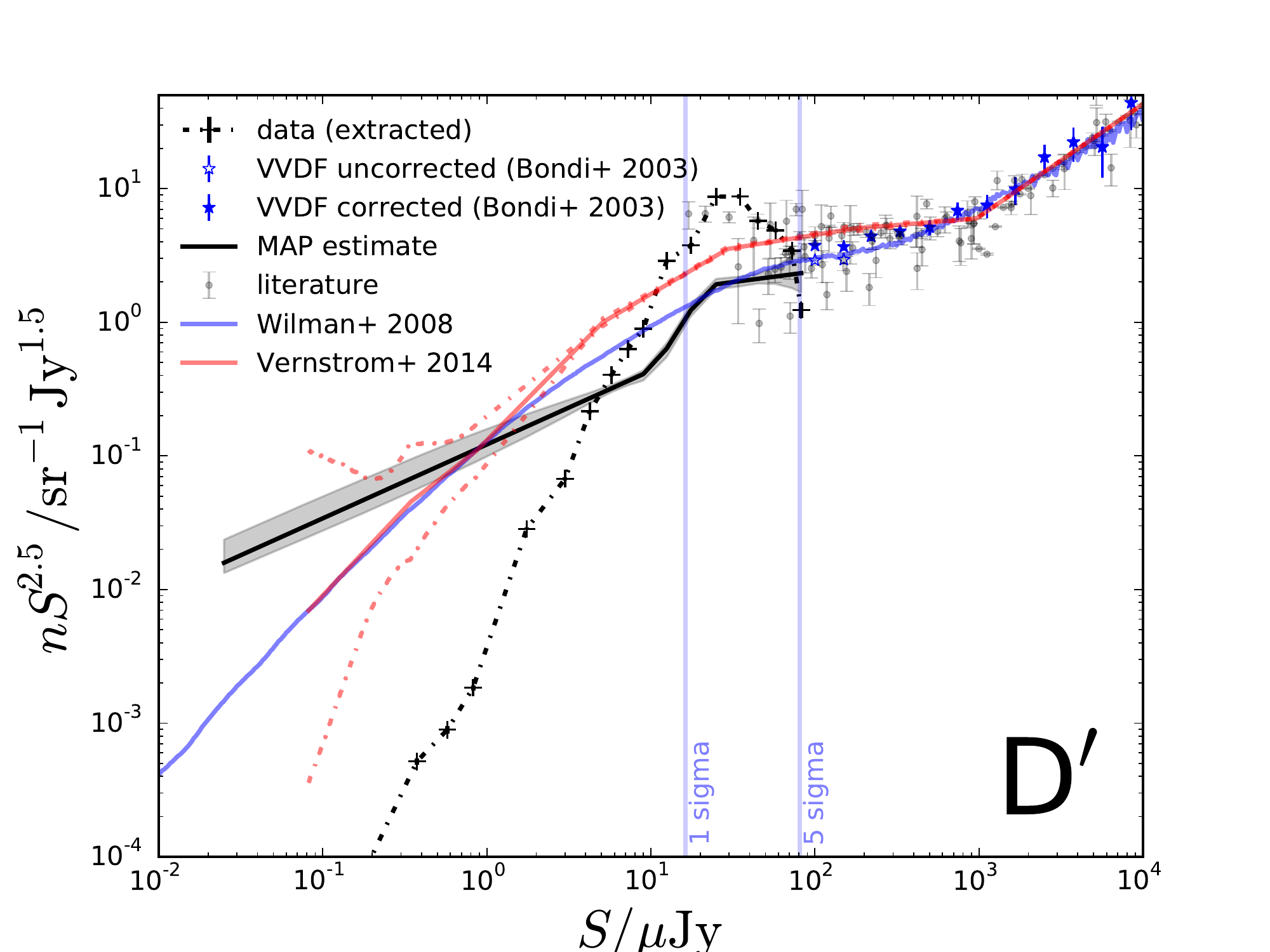}
\caption[]{Reconstructions for all galaxies in the VIDEO sample. The
  model fitted is indicated bottom right, the preferred one being
  model B\arcmin.
The underlying full SKADS
  \citep{wilman2008} counts are shown in blue, and the only other
  counts at these depths \citep{vernstrom2014} are marked in red, with
  the 68~per~cent confidence interval in dashed red. The filled and
  empty blue stars respectively represent the \cite{bondi2003}
  corrected and uncorrected counts for the VLA $>5\sigma$ sources.
  Literature values are taken from the review by
  \cite{dezotti-review}.
The real, noisy data are shown as a black dashed line. The black line and
  grey shaded region show the maximum \textit{a posteriori} (MAP) and
  68-per-cent confidence interval of the distributions of models reconstructed from
  every sample from the posterior. The
  vertical blue lines are at 1$\sigma$ and 5$\sigma$.
\label{fig:recon-results-all2}}
\end{figure*}

\begin{figure*}
\centering
\includegraphics[width=1.0\textwidth,origin=br,angle=0]{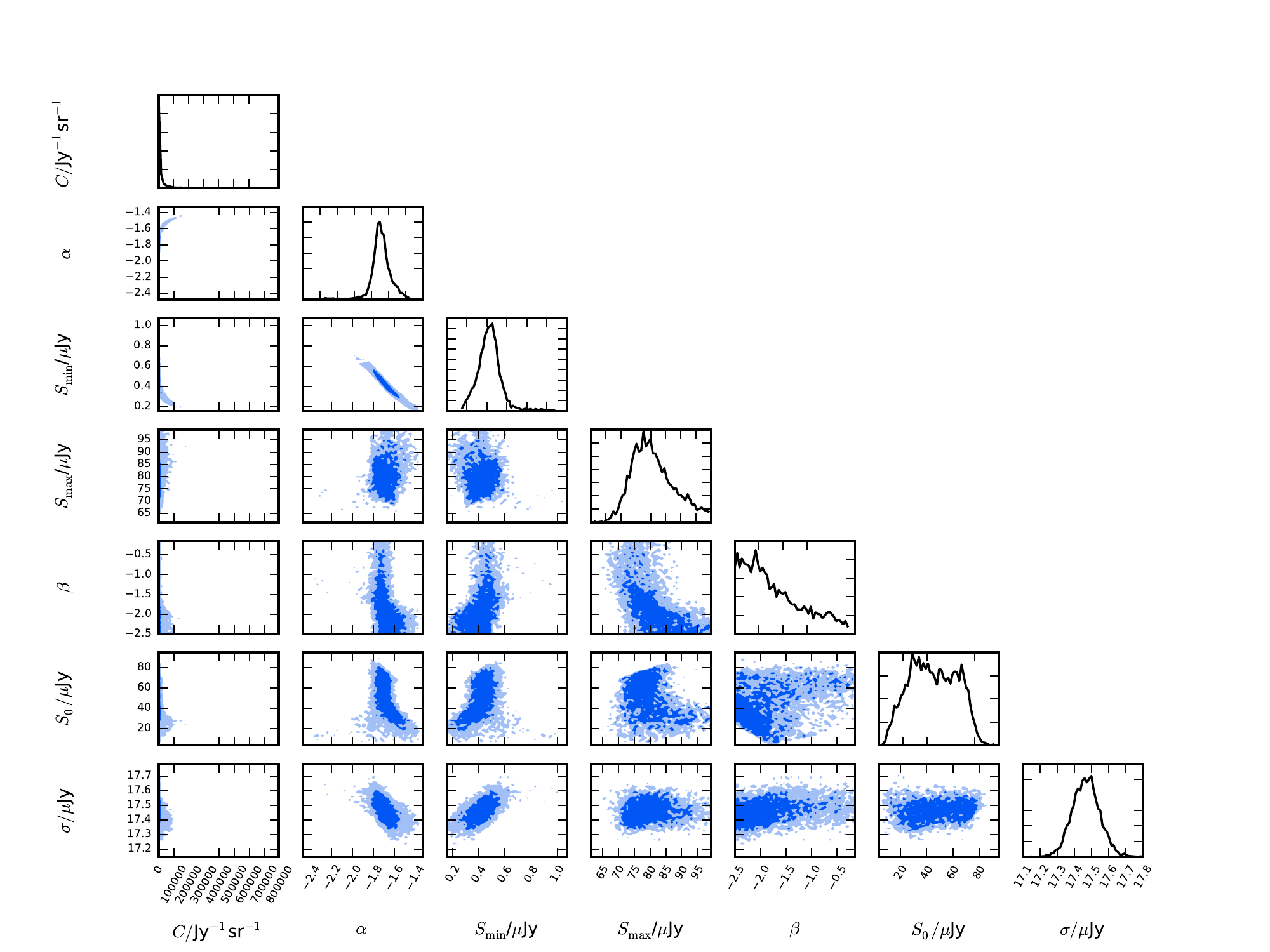}
\caption{Posterior probability distribution for the preferred
  three-break model~B\arcmin\ fitted to the VIDEO data. 68 and
  95~per~cent confidence limits are shown as dark and light blue
  shaded regions respectively.\label{fig:video-tri2}}
\end{figure*}

The reconstructions for the preferred models (from a choice of A, B, C
and~D for SKADS, and from all eight models for VIDEO) fitted to the
resulting flux histograms are shown in
Figure~\ref{fig:recon-results-random}. The comparison is revealing.
For the SKADS extraction, the inferred counts are flat in euclidean
space, with a very slight ramp at the bright end, with small
uncertainties caused by a high signal-to-noise measurement of the
average background flux. The VIDEO counts have a ramp at 1--5$\sigma$
due to confusion from unmasked sources in that regime (consider, for
example, flux boosting due to an unmasked 4.5$\sigma$ source).
However, the overall amplitude is considerably suppressed relative to
the signal from the true positions (note the logarithmic scale), even
including the ramp. Moreover, the amplitude is suppressed by a factor
of $\approx$20 compared to the SKADS prediction.
We conclude that our results are not contaminated by confusion at any
notable level.

\footnotetext{These are available from\\
  \texttt{http://web.oapd.inaf.it/rstools/srccnt/srccnt\_tables.html}}

\begin{table}
\centering
\caption{Evidences $\Delta\mathcal{Z}$ for models A--D
  applied to the VIDEO data, relative to model~A. It appears that the preferred model
  at this stage is D, but see Table~\ref{table:data-evs2}.
  \label{table:data-evs}}
\begin{tabular}{lcc}
\hline
Model \T\B & $\Delta\log_{\mathrm{e}}\mathcal{Z}$ & Odds ratio $\mathcal{Z}/\mathcal{Z}_A$\\
\hline
A \T & 0.0 & 1 \\
B & 7.44$\pm$0.22 & $10^{3.2}$:1 \\
C & 21.24$\pm$0.24 & $10^{9.2}$:1 \\
D \B & \textbf{27.65}$\mathbf{\pm}$\textbf{0.24} & $10^{12.0}$:1 \\
\hline
\end{tabular}
\end{table}

\begin{table}
\centering
\caption{Evidences $\Delta\mathcal{Z}$ for models A\arcmin--D\arcmin\
  applied to the VIDEO data, relative to model A. The preferred model
  is now (just) B\arcmin.
  \label{table:data-evs2}}
\begin{tabular}{lcc}
\hline
Model \T\B & $\Delta\log_{\mathrm{e}}\mathcal{Z}$ & Odds ratio $\mathcal{Z}/\mathcal{Z}_A$\\
\hline
A \T & 0.0 & 1 \\
A\arcmin & 152.51$\pm$0.21 & $10^{66.23}$:1 \\
B\arcmin & \textbf{152.86}$\mathbf{\pm}$\textbf{0.21} & $10^{66.39}$:1 \\
C\arcmin & 152.70$\pm$0.21 & $10^{66.32}$:1 \\
D\arcmin \B & 152.48$\pm$0.21 & $10^{66.22}$:1 \\
\hline
\end{tabular}
\end{table}

\subsection{Subdivision by galaxy type}
\label{sec:results:mod-best}

\begin{figure*}
\centering
\includegraphics[width=8.5cm,origin=br,angle=0]{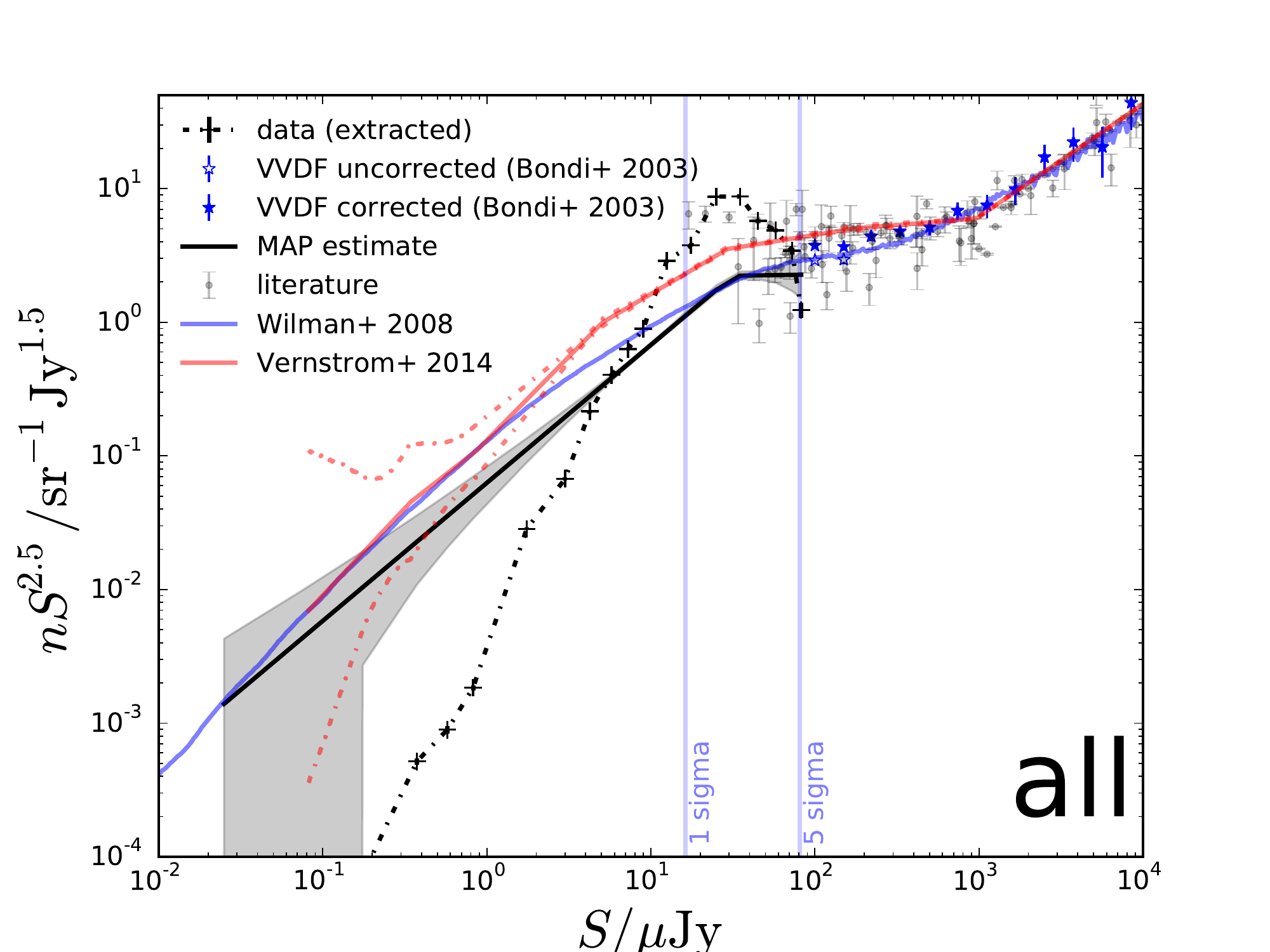}
\includegraphics[width=8.5cm,origin=br,angle=0]{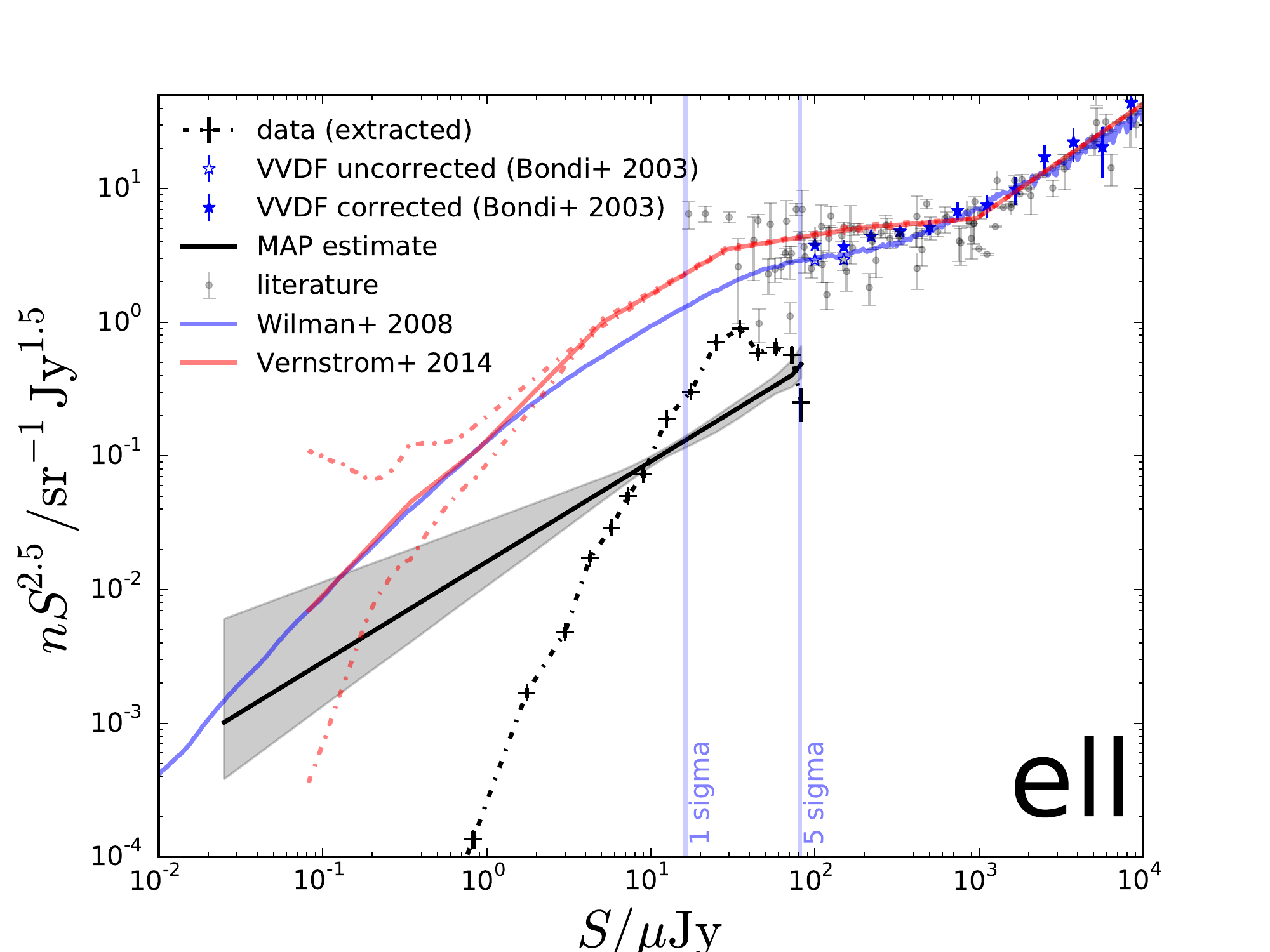}\\
\includegraphics[width=8.5cm,origin=br,angle=0]{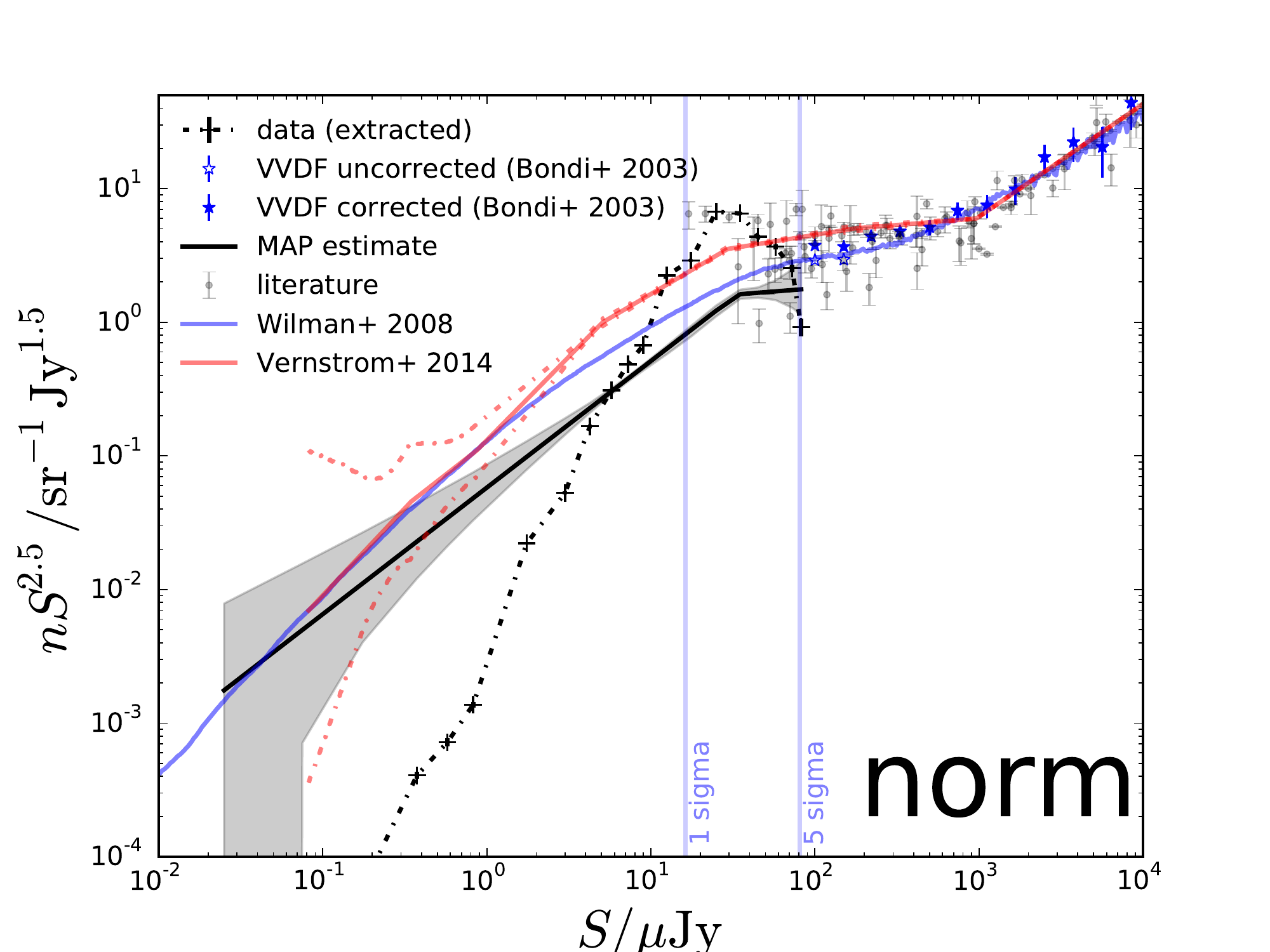}
\includegraphics[width=8.5cm,origin=br,angle=0]{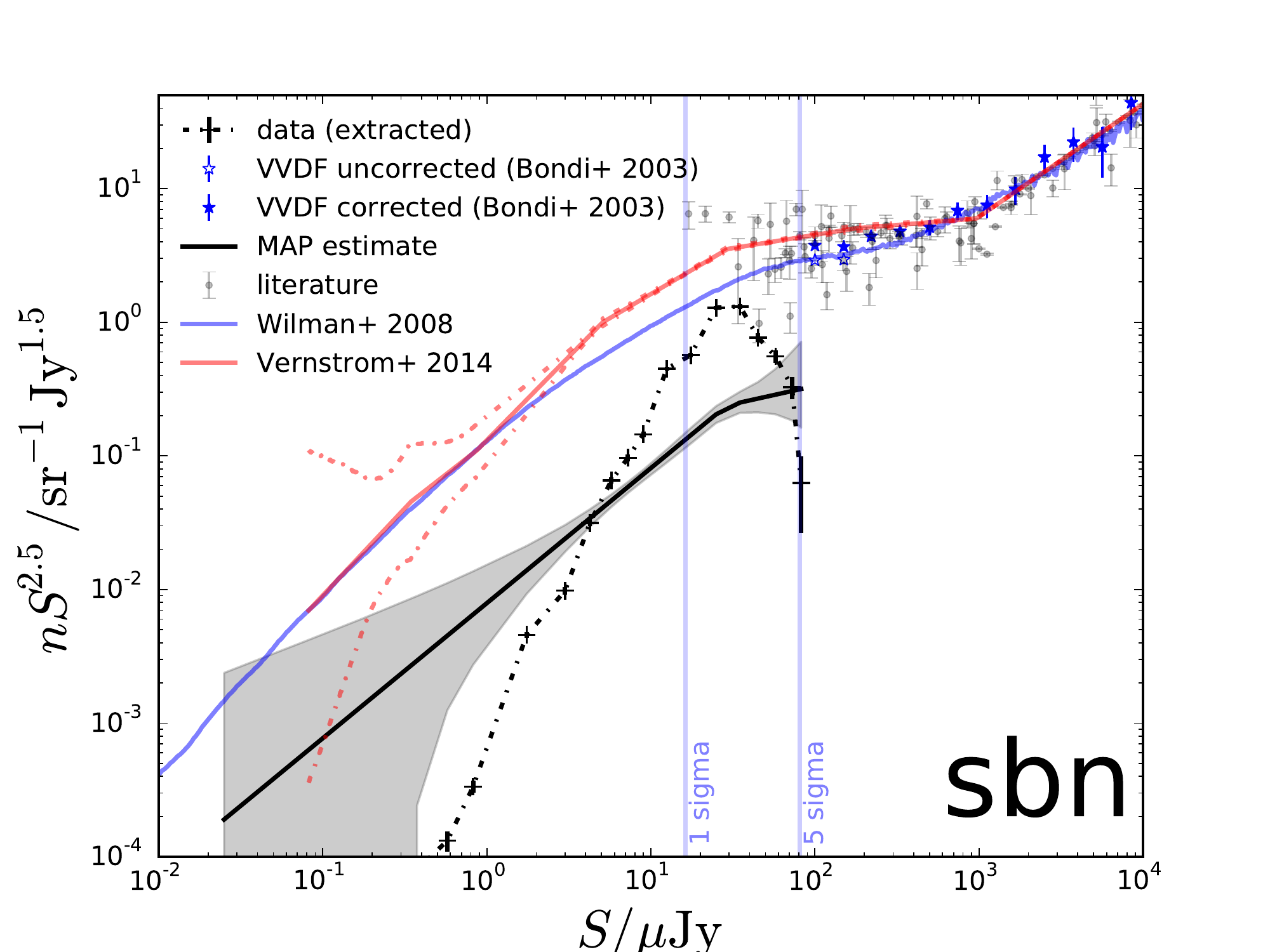}
\caption[]{Reconstructions for galaxies in the VIDEO sample, by galaxy
  subsample (all, elliptical, normal and starburst).
 The underlying full SKADS
  \citep{wilman2008} counts are shown in blue, and the only other
  counts at these depths \citep{vernstrom2014} are marked in red, with
  the 68~per~cent confidence interval in dashed red. The filled and
  empty blue stars respectively represent the \cite{bondi2003}
  corrected and uncorrected counts for the VLA $>5\sigma$ sources.
  Literature values are taken from the review by
  \cite{dezotti-review}.
The real, noisy data are shown as a black dashed line. The black line and
  grey shaded region show the maximum \textit{a posteriori} (MAP) and
  68-per-cent confidence interval of the distributions of models reconstructed from
  every sample from the posterior. The
  vertical blue lines are at 1$\sigma$ and 5$\sigma$.
\label{fig:mod-best}}
\end{figure*}

\noindent
We subdivided the galaxy sample using the criteria of
\cite{zwart2014}, which are based on the best-fitting spectral-energy
distribution (to the 10-band photometry), which \citeauthor{zwart2014}
argue to be a better discriminator than a colour-colour diagram. The
three subsample classifications are (i) ellipticals, (ii) normal (Sbc,
Scd, low-mass irregular) galaxies and (iii) starburst galaxies.  We
fitted models A\arcmin--D\arcmin\ to the three subsamples from this
scheme in order to investigate the relative contributions to the
source counts of the different populations. The results are summarized
in Table~\ref{table:mod-best} and Figure~\ref{fig:mod-best}. The
counts are dominated by the more numerous `normal' galaxies, whose
counts also exhibit the (full sample's) flat behaviour down to
$\simeq 20 \mu$Jy. The counts of the ellipticals do not flatten at the
bright end, but fall at a shallower rate, one that is consistent with
the very bright ($>1$\,mJy) slope, as expected for such AGN hosts. The
starburst counts are not as flat as the overall trend as far as
20$\mu$Jy, but do, like the other subsamples, steepen below that
flux. Hence, as expected, the source counts are dominated by normal
`spiral galaxies' at these fluxes, with starburst galaxies and AGN
making little contribution (our selection excludes QSOs).

\begin{table}
\centering
\caption{Summary of analysis for the VIDEO galaxy-type subsamples.\label{table:mod-best}}
\begin{tabular}{llccc}
\hline
Type \T\B & Galaxies & Flux range / $\mu$Jy & Bins & Preferred model \\
\hline
All \T & 71,418 & $-$108--85 & 41 & B\arcmin\\
Elliptical & 5351 & $-$67--85 & 38 & B\arcmin\\
Normal & 54,879 & $-$108--85 & 41 & C\arcmin\\
Starburst \B & 11,187 & $-$69--85 & 38 & B\arcmin \\
\hline
\end{tabular}
\end{table}

\section{Conclusions}
\label{sec:conclude}

\begin{enumerate}

\item We have cast `sub-threshold stacking' in a fully bayesian
  framework for the first time, including the ability to calculate the
  evidence for the purposes of model selection.

\item As well as the bayesian evidence, our framework reveals the
  exploration of full posterior probability distributions, showing
  explicitly any degeneracies and/or correlations. We note that
  marginalized parameter estimates are incorrect if such degeneracies
  are present as in these cases.

\item We applied the algorithm to the SKADS simulations. When run on a
  SKADS catalogue, we were able to reconstruct the counts successfully
  down to sub-$\mu$Jy levels, i.e.~the $5\sigma_n/\sqrt{N=375000}$
  level. This also holds true when applied to data (below).

\item We used the SKADS catalogue to simulate a VIDEO--VLA-like radio
  map from which we extracted fluxes at the positions of SKADS sources on
  which to run the algorithm. We showed that confusion biased the
  counts high unless we artifically assumed no radio emission below
  1\,$\mu$Jy. This needs to be accounted for in analysis of data from
  relatively-low-resolution SKA pathfinders such as ASKAP.

\item We applied the algorithm to VLA data stacked at VIDEO
  positions. A power-law model (D) with three breaks (the two at
  3\,$\mu$Jy and 20\,$\mu$Jy being the predominant ones) is preferred
  when the modal map noise of 16.2\,$\mu$Jy is assumed.

\item When the map noise is varied as a free parameter, a power-law
  model (B\arcmin) with a single break is overwhelmingly preferred,
  and we adopt model B\arcmin\ (over the fixed-noise model~D) as our
  final result. The inferred counts can be summarized as flat to
  $\approx 40\mu$Jy, then falling with a slope of $-$1.7 below that
  flux. The map noise estimated via is this route is, rather,
  $17.5\pm 0.08\mu$Jy.

\item While one would not expect them to match \textit{a priori}, we
  interpret the slight deficit of counts below 20\,$\mu$Jy relative to
  the results of \cite{wilman2008} and \cite{vernstrom2014} as
  indicative of a fraction of the radio emission coming from galaxies
  fainter than the flux limit of the VIDEO catalogue, be they
  lower-redshift faint galaxies with some ongoing star formation, or
  higher-redshift moderately bright galaxies.
  Like those works, our results are not consistent with the ARCADE2
  results or indeed those from the work of \cite{om2008}.

\item We have subdivided the VIDEO sample into ellipticals, normal and
  starburst galaxies, and fitted source count models to the binned
  flux distributions, finding that the counts are dominated by normal
  `spiral galaxies' at these fluxes, with little contribution from
  starburst galaxies or AGN.

\item We note the usefulness and wide applicability of our algorithm,
  and look forward to its employment in future radio surveys such as
  those with MeerKAT \citep{meerkat-surveys}, LOFAR \citep{lofar},
  ASKAP \citep{emu} and SKA \citep{sparcs,ska-continuum-overview}.

\end{enumerate}

\subsection{Future work}
\label{sec:conclude:future}

The radio luminosity, solely a function of source flux and redshift,
can be used to infer star-formation rates, and, using the available
stellar-mass estimates, specific star-formation rates
(cf.~\citealt{zwart2014}). It will be straightforward to extend our
algorithm to the derivation of luminosity functions
(Malefahlo~et~al.~in~prep.), and their evolution with redshift,
building on our work and that of \cite{R+B2013}. It will be
interesting to compare estimates for these quantities with those
derived from traditional stacking methods. As hinted at earlier, one
would like to undertake a joint analysis of the counts of confusing
sources below the thermal-noise limit together with those measured in
this work, and our framework permits this too.  Finally, the algorithm
can readily be applied to any survey data map that has an auxiliary
catalogue of greater depth and resolution, e.g.~VLA-COSMOS, 10C
(Whittam~et~al.~in~prep.), BLAST or \textit{Herschel}-ATLAS, though in
some low-resolution cases modification may be needed in order to
account for a $P(D)$-style confusion contribution (high likelihood of
beam occupancy $>$1).

\section*{Acknowledgments}
\label{sec:acknowledgments}

JZ gratefully acknowledges a South Africa National Research Foundation
Square Kilometre Array Research Fellowship. MGS acknowledges support
from the South African Square Kilometre Array Project, the South
African National Research Foundation and FCT-Portugal under grant
PTDC/FIS-AST/2194/2012. MJJ is grateful to the South African Square
Kilometre Array Project for financial support. We thank Farhan Feroz,
Mat Smith, Russell Johnston, Jasper Wall, Tessa Vernstrom, Ian Heywood
and Johannes Buchner for useful discussions. The authors thankfully
acknowledge the computer resources, technical expertise and assistance
provided by CENTRA/IST. We especially thank Sergio Almeida for
valuable computing support. Computations were performed at the cluster
``Baltasar-Sete-Sois'' and supported by the DyBHo-256667 ERC Starting
Grant. This work is based on data products from observations made with
ESO Telescopes at the La Silla or Paranal Observatories under ESO
programme ID~179.A-2006.

\bibliography{bayestack}\label{lastpage}
\bibliographystyle{astron}
\bsp

\onecolumn

\end{document}